\documentclass[trackchanges,twocolumn]{aastex701}

\newcommand{\e}[1]{\times 10^{#1}}
\usepackage{amsmath}

\begin{document}

\title{An early look at how gas giants shape small planet bulk compositions}

\author[orcid=0009-0006-2841-8101]{Joseph Y. Tang}
\affiliation{Department of Astronomy, Columbia University, New York, NY, 10027, USA}
\email{jyt2124@columbia.edu}

\author[orcid=0000-0002-6076-5967, gname=Marta, sname=Bryan]{Marta L. Bryan} 
\affiliation{Department of Astronomy \& Astrophysics, The Pennsylvania State University, 201 Old Main, University Park, PA 16802-1589, USA}
\email{martalbryan@psu.edu}

\author[orcid=0000-0002-1228-9820,gname=Eve, sname=Lee]{Eve J.~Lee}
\affiliation{Department of Astronomy \& Astrophysics, University of California, San Diego, La Jolla, CA 92093-0424, USA}
\email{evelee@ucsd.edu}

\begin{abstract}

Gas giants may shape the reservoir of solids and gas in the inner disk in which the small planets assemble. To test this possibility, we collect a sample of 43 exoplanetary systems containing 68 inner small planets (ISP) with both measured masses (1--20 M$_{\oplus}$) and radii (1--4 R$_{\oplus}$). After correcting for heterogeneous individual system sensitivities to distant gas giants, we calculate the gas giant occurrence rate in ISP systems P(GG$|$ISP) as a function of inner small planet density, envelope mass fraction (EMF), and core mass. While we find no significant difference between P(GG$|$ISP) given high/low small planet density, EMF, or core mass, we see hints of a trend when only looking at the metal-rich systems. Despite the substantial limitations due to small sample sizes, we find that gas giants in metal-rich systems are preferentially found with lower density planets with similar core masses. We find consistent {\it hints of} trends using larger samples of inner planets with measured radii divided across the radius valley or with measured masses divided across 10 $M_\oplus$. Our result is consistent with more metal-enriched disks catalyzing rapid core assembly and kickstarting the gas accretion early, while the muted difference in the outer giant occurrence rate with respect to core mass may indicate contamination by post-formation photoevaporation.

\end{abstract}

\section{Introduction} \label{sec:intro}

Small planets with masses 1--20 M$_{\oplus}$ and radii 1--4 R$_{\oplus}$ are the most common type of detected exoplanet, with an estimated occurrence of $\sim30\%$ orbiting sun-like stars and an even higher frequency around M-dwarf hosts \citep[e.g.][]{Zhu2018,Ment2023,Pinamonti2022}. Recent work has shown that gas giant analogs to Jupiter are common in small planet systems, and that this positive correlation between these two planet populations is a strong function of host star metallicity \citep{BryanML2024a,BryanML2025,Zhu2024,Bonomo2025}. Metal-rich FGK stars with inner small planets are more likely to host outer gas giants than metal-rich field FGK stars. This enhanced inner-outer planet correlation in metal-rich systems points to the critical role that metallicity, as a proxy for disk solid content, plays in shaping architectures of planetary systems \citep{ChachanY2023,BryanML2024,BryanML2025}.  

A fundamental open question in this quest to constrain the gas giant/small planet connection is: do gas giants shape the bulk composition of inner small planets? Previous theoretical works have argued that gas giants can open gaps in their natal disks and restrict the inward flow of planetary building blocks \citep[e.g.,][]{Morbidelli15,Lambrechts19} or kick volatile-rich planetesimals inwards \citep[e.g.,][]{WalshKJ2011,BatyginK2015} that can dynamically interact with inner planets in destructive ways. 
The observed inner-outer planet correlation rules out the previously imagined harmful role that the outer giant plays in the creation or the stability of the inner planets. So, does the outer planet play an active role in shaping the reservoir of solids and gas in the inner disk where the small planets assemble, or is it a passive bystander---a byproduct of initial disk properties? 

Observational constraints on the composition of small worlds are challenging. The James Webb Space Telescope (JWST) presents an exciting opportunity to target atmospheric compositions of small planets, but these measurements are both time-intensive and frequently foiled by degeneracies between clouds and metal-rich atmospheric compositions \citep[e.g.][]{Ahrer2025,Ohno2025,Rochon2025}. Gathering enough small planet systems with atmospheric composition constraints from JWST to look for correlations between composition and the presence of a gas giant will take time. Furthermore, it remains unclear how much the the upper atmosphere can reveal about the composition of the deeper interior.

Here we adopt a timely alternative strategy. We aim to explore the connection between inner small planet bulk composition and the presence of outer gas giant companions using a population of small planets with both mass \textit{and} radius measurements. In addition to straightforward small planet density constraints, we leverage modeling work from \citet{LopezED2014} to back out estimates of core masses and envelope mass fractions (EMFs) for these small planets. 

We discuss our inner small planet sample in Section~\ref{sec:sample collection} and how we account for sample biases dominated by different system sensitivities to distant gas giants in Section~\ref{sec:injection recovery}. Section~\ref{sec:sanity check} describes a sanity check for this new sample before Section~\ref{sec:correlation exploration} presents our results quantifying frequencies of gas giants in small planet systems as a function of planet density, core mass, and EMF. We expand our sample size in Section \ref{sec: no diff valley} by looking at planets with radius \textit{or} mass measurements and quantify occurrences for planets falling above/below the radius valley and at high/low masses. Finally, Section~\ref{sec:discussion} discusses the implications of our results in the context of relevant planet formation theory. We conclude in Section~\ref{sec:conclusion}.

\section{A sample of small planets with both mass and radius measurements} \label{sec:sample collection}

Our goal in this study is to determine the occurrence rate of outer gas giants in systems with inner small planets that have both mass and radius measurements. To achieve this, we design our selection criteria as follows: (1) a system must have publicly available RV dataset(s); (2) RV datasets must have an observational time baseline $>$1 year and at least 20 data points, yielding minimum system sensitivities to outer gas giants; (3) systems must have at least one small planet with a measured mass (1--20 M$_{\oplus}$) and radius (1--4 R$_{\oplus}$); and (4) host star masses must be M$_\star$$>$0.6 M$_{\odot}$ (excluding M-dwarfs). We exclude M-dwarf systems because the frequency of outer gas giants to small planet systems is markedly lower than that of FGK systems \citep{BryanML2025}. Applying these criteria yields a sample of 43 FGK star systems hosting 68 inner small planets that have both mass and radius measurements. The sample assembled here is a subset of the 184 small planet system sample presented in \citet{BryanML2024}. 

\begin{figure}
    \centering
    \includegraphics[width=1\linewidth]{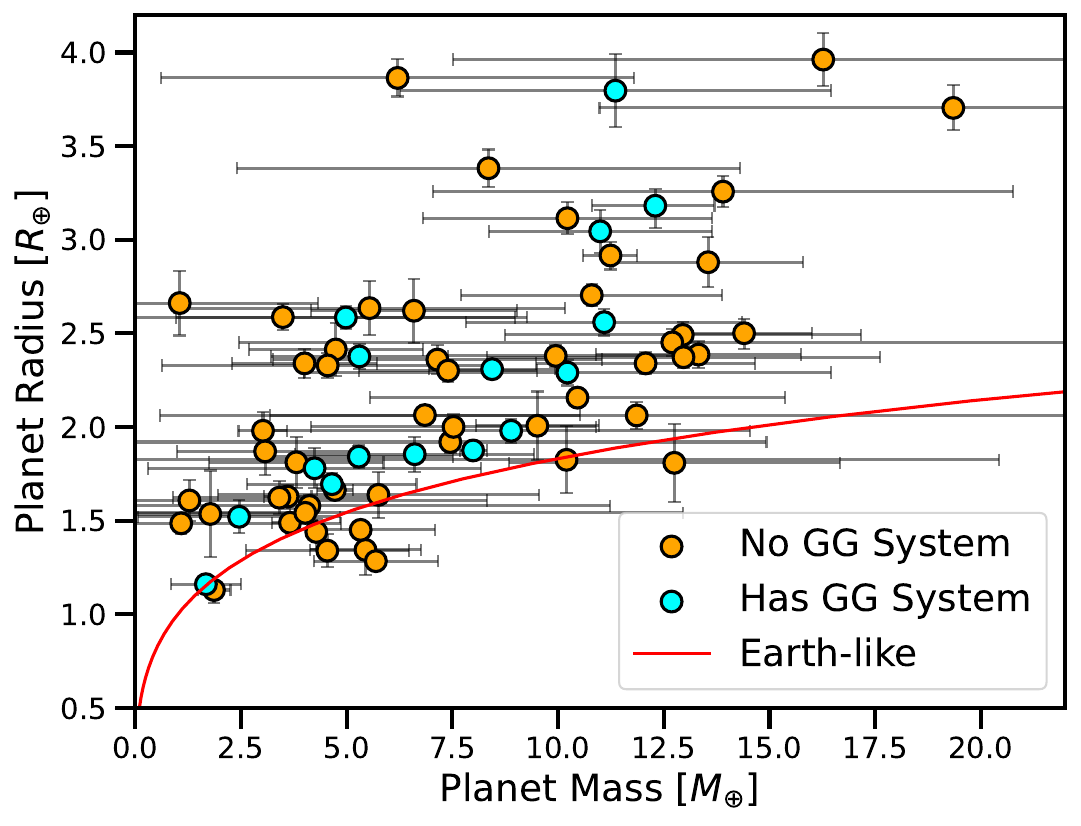}
    \includegraphics[width=1\linewidth]{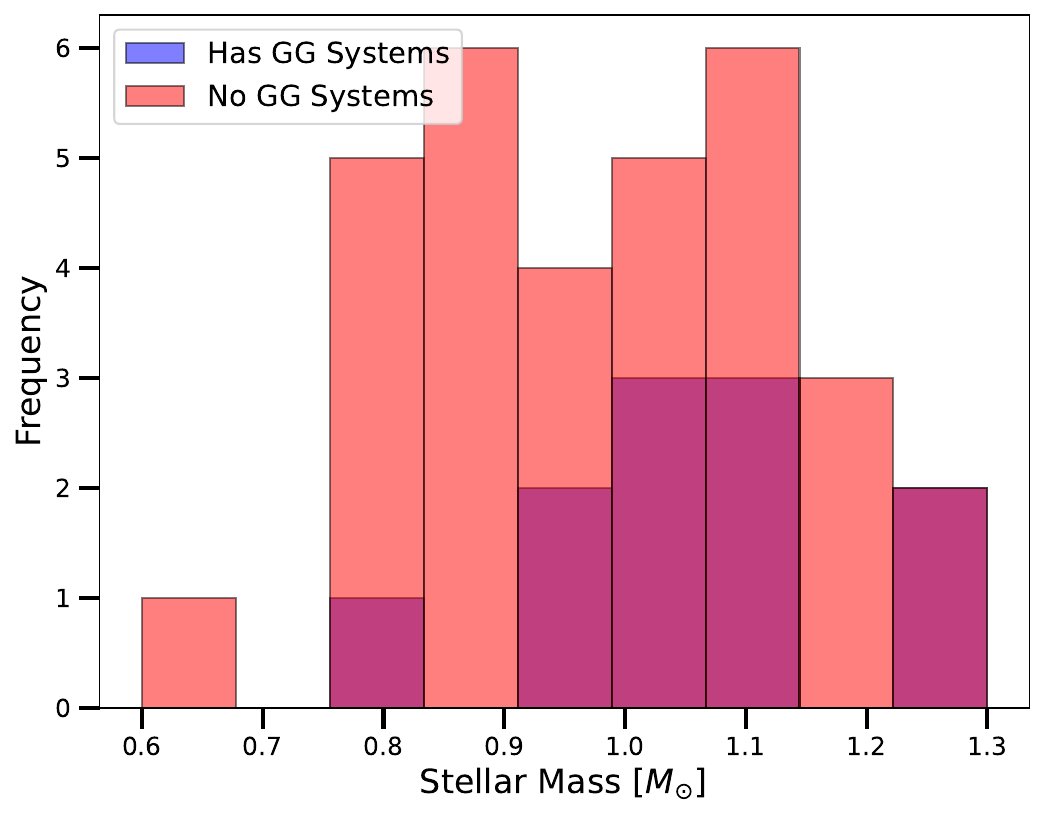}
    \includegraphics[width=1\linewidth]{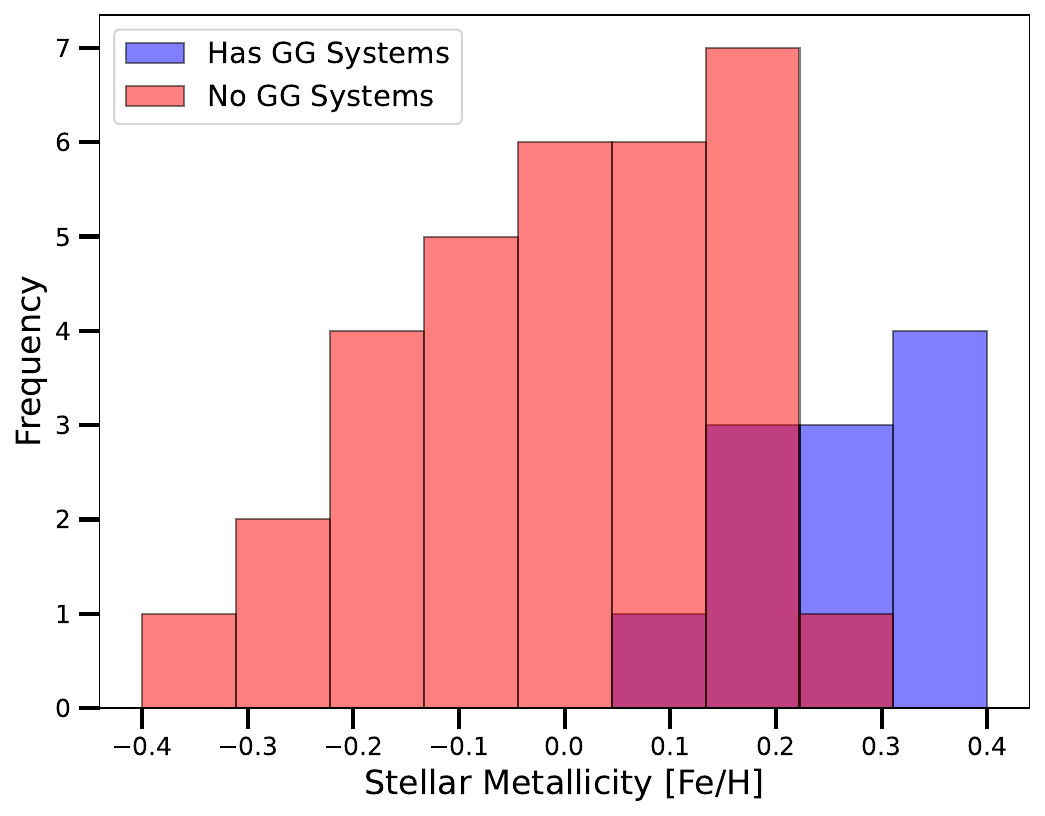}
    
    \caption{\textbf{Top:} Mass-radius diagram for small planets included in our sample. Small planets in systems with and without gas giants are plotted in blue and orange, respectively. We include a density curve for an Earth-like planet in red from \cite{ZengL2019}. \textbf{Middle:} Stellar mass distributions for our sample, split into systems with gas giants (blue) and without gas giants (red). \textbf{Bottom:} Stellar metallicity distributions for our sample, again separated into systems with (blue) and without (red) gas giants.}
    \label{fig:scatter and hist}
\end{figure}

In our sample of 43 small planet systems (Table~\ref{tab:systems}), 27 of them have metal-rich ([Fe/H]$>$0) host stars, while the remaining 16 systems have metal-poor ([Fe/H]$\leq$0) stars. 11 of the 43 small planet systems have detected gas giants (GG; 0.5--20 M$_{\rm Jup}$), with a total of 13 cold, outer GG's (1--10 AU), and 3 warm Jupiters (0.1--1 AU). In all three warm Jupiter systems (KOI-142, Kepler-65, and 55 Cnc), the small planets lie interior to the warm Jupiter. We show the mass-radius diagram for the small planets included in our sample in the top panel of Figure~\ref {fig:scatter and hist}, distinguishing between small planets in systems with and without detected outer gas giants. Stellar properties of our sample are presented in the bottom two panels of this figure, showing consistent stellar mass distributions between systems with and without gas giants, and enhanced stellar metallicities for systems with gas giants. In fact, the 11 systems with gas giants all have metal-rich host stars.

\section{Quantifying individual system sensitivities to distant gas giants} \label{sec:injection recovery}
The RV datasets for each individual system are heterogeneous, with different instruments, precisions, time baselines, number of datapoints, and observing cadences. All of these impact the sensitivity of a given RV dataset to the presence of an outer gas giant. To quantify each system's sensitivity to distant gas giants, we calculate completeness maps in planet mass and semi-major axis space for individual systems. 

Starting from a $50\times50$ logarithmically spaced grid in mass and semi-major axis spanning 0.3--30 M$_{\rm Jup}$ and 0.3--30 AU, for each system, we inject 50 simulated planets in a given gridbox. For each planet, we randomly draw that planet's mass and semi-major axis from the gridbox, the planet's eccentricity from a $\beta$ distribution, inclination $i$ from a uniform distribution in $\cos i$, and remaining orbital elements from uniform distributions. This yields an orbital solution for each simulated planet.

Using these orbital solutions and the observational epochs of the actual RV dataset for a given system, we generate simulated RVs at every epoch for each simulated planet. Associated uncertainties on those RVs are taken from the actual dataset, shuffled, and assigned to the simulated RVs.

\begin{figure}[!htb]
    \centering
    \includegraphics[width=1\linewidth]{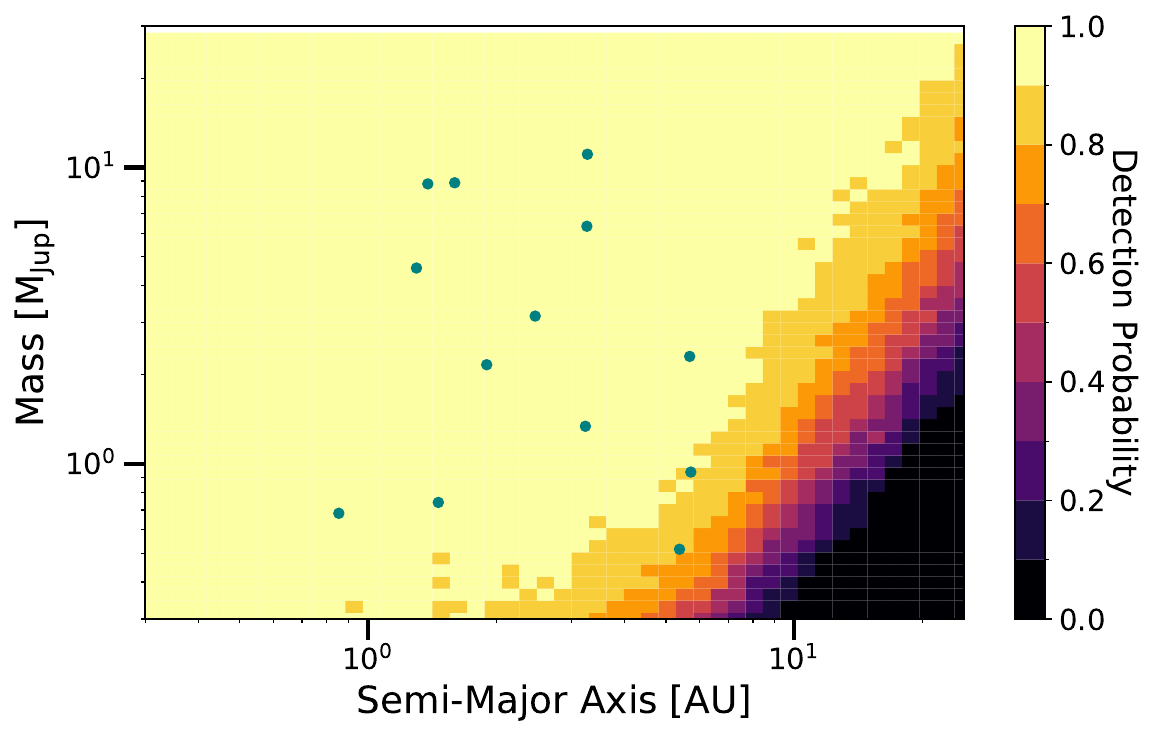}
    \caption{The median completeness map for all 43 systems. The 13 detected cold gas giants (0.5--20 M$_{\rm Jup}$, 1--10 AU) are plotted as blue dots. Note that all gas giants lie in a region of high sensitivity parameter space for our sample.}
    \label{fig:avg completeness maps}
\end{figure}

Given the simulated RV dataset, we want to determine whether this simulated planet would be ``detected'' or ``not detected''. We fit two different models to the simulated RVs; a one-planet orbital solution, and a flat line. These model fits are compared using their respective Bayesian information criterion (BIC) values. If the one-planet solution is preferred by a factor of 10 or more than the flat line, then we consider the simulated planet ``detected''. If the one-planet solution is preferred by less than a factor of 10 or not preferred at all, the simulated planet was ``not detected''. We applied this process to all 50 simulated planets/gridbox over the entire 50$\times$50 grid of planet masses and semi-major axes for each system. Each grid cell of these sensitivity maps thus contains the probability, between 0 and 1, that a gas giant in that mass and semi-major axis space would be detected given the observational RV dataset. Figure~\ref {fig:avg completeness maps} shows the median completeness map for all 43 systems in our sample, with the 13 detected gas giants overplotted. Note that all gas giants lie in a region of high sensitivity parameter space for our sample.

\begin{figure}
    \centering
    \includegraphics[width=0.93\linewidth]{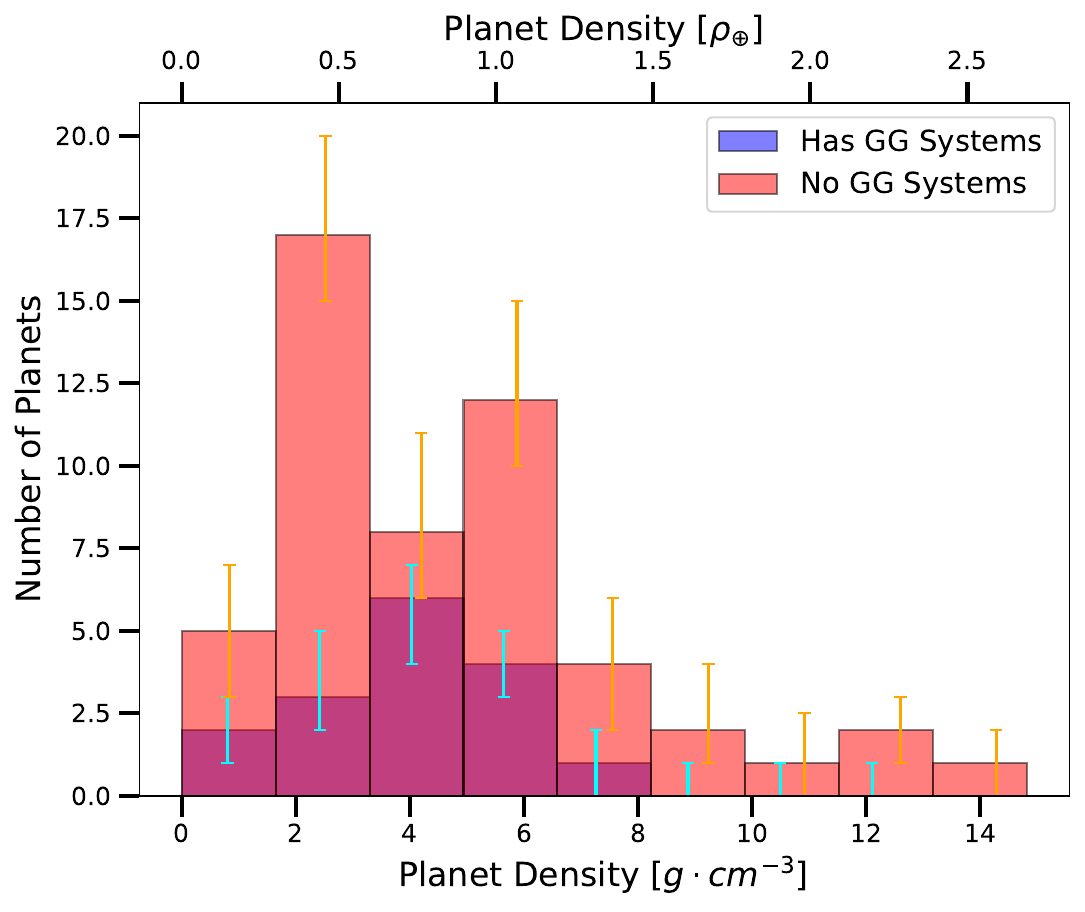}
    \includegraphics[width=0.93\linewidth]{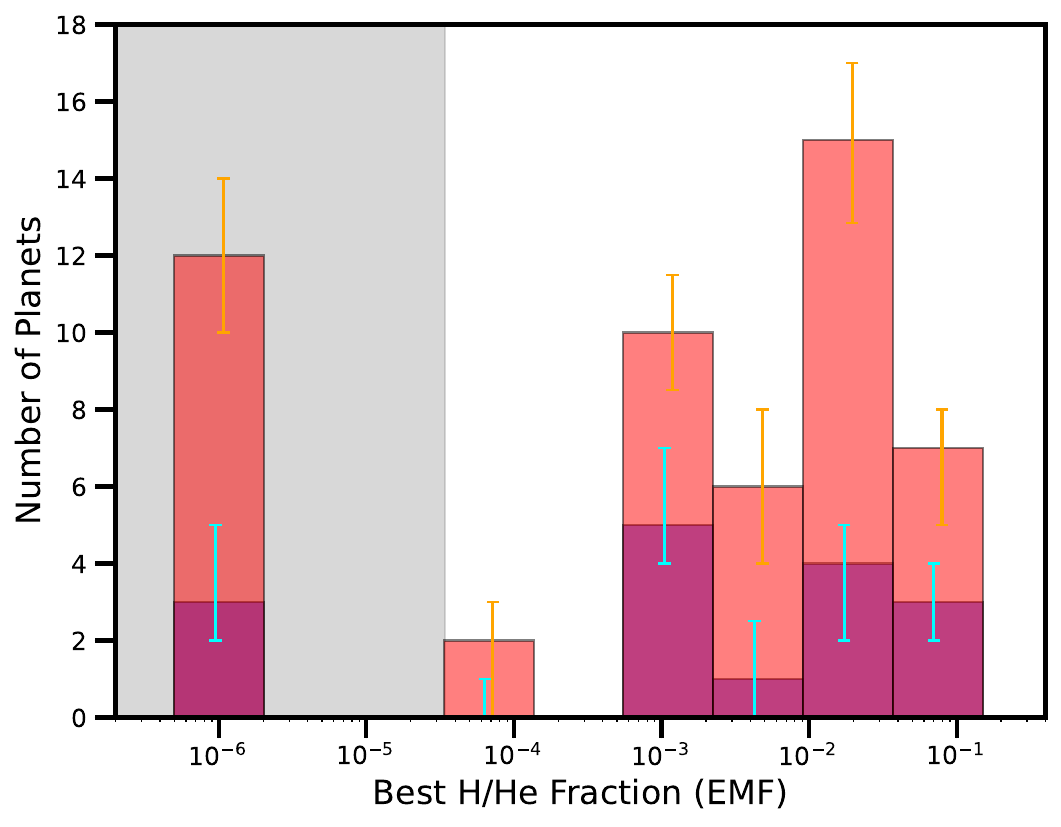}
    \includegraphics[width=0.93\linewidth]{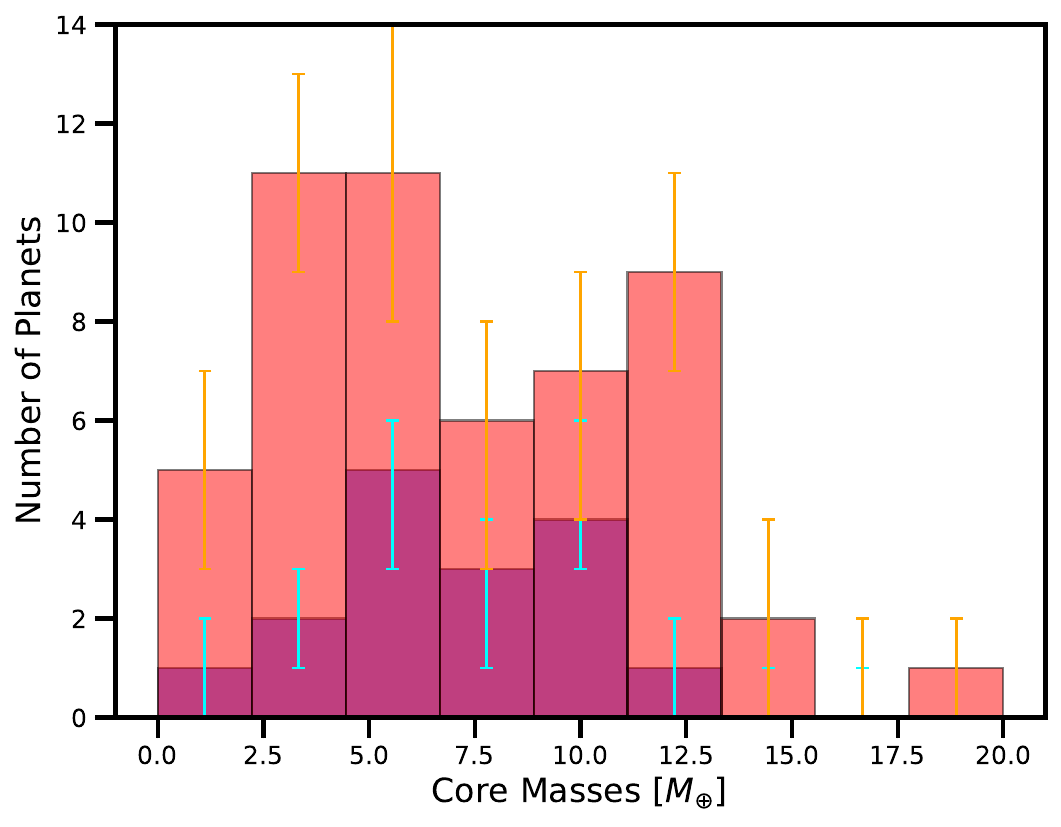}
    \caption{A histogram of planet densities \textbf{(top)}, envelope mass fractions (EMFs) \textbf{(middle)}, and core masses \textbf{(bottom)} for our sample of small planets. Blue and red colors represent systems with and without gas giants, respectively. The gray zone in the middle panel represents the empty region below the lowest curve in \cite{LopezED2014} where no EMF is provided. Instead, all H/He fractions are set to $10^{-6}$ to represent an Earth-like EMF.}
    \label{fig:pl hists}
\end{figure}

\section{A sanity check to confirm occurrence rate P(GG$|$ISP) for our sample} \label{sec:sanity check}

With our sample of 43 small planet systems that have both measured masses and radii, we can explore the question of whether the presence or absence of an outer gas giant correlates with inner small planet density, envelope mass fraction, and core mass. Before tackling these questions, as a sanity check we first calculate P(GG$|$ISP) following \citet{BryanML2024, BryanML2025} to confirm consistency with the larger samples previously published. 

We start with the general binomial distribution:

\begin{equation}
    f(x;a,b) = \frac{1}{B(a,b)}x^{a-1}(1-x)^{b-1}
\end{equation}
where $B$ is the beta distribution, $a = n_{\text{det}} - 1$, and $b = n_{\text{eff}} - n_{\text{det}} + 1$. Here $n_{\rm det}$ is the number of systems with detected gas giants, and $n_{\text{eff}}$ is the effective total number of systems in our sample modulated by individual system sensitivities. 

Splitting our sample of 43 systems by host star metallicity, we calculate $\mathrm{P(GG|ISP, [Fe/H]}>0) = 0.455^{+0.10}_{-0.094}$, where uncertainties correspond to the 1$\sigma$ highest probability density interval (HPDI), and $\mathrm{P(GG|ISP, [Fe/H]} \leq 0)<0.074$, the 1$\sigma$ upper limit. Comparing these occurrence rates with those obtained in \citet{BryanML2024}---namely P(GG$|$ISP, [Fe/H] $>$ 0) $=0.280^{+0.049}_{-0.046}$ and P(GG$|$ISP, [Fe/H] $\leq0$) $=0.045^{+0.026}_{-0.019}$---we find that corresponding positive and negative frequencies are consistent to $<1\sigma$.

\section{Do gas giants impact inner planet densities, envelope mass fractions, and core masses?} \label{sec:correlation exploration}

We now assess the effect of outer gas giants on the densities, envelope mass fractions (EMF), and core masses of the inner planets. We start by using the measured small planet masses and radii in conjunction with models from \citet{LopezED2014} to compute these three bulk composition characterizations.

To reference below, the distributions of these aforementioned inner planet properties are shown in Figure~\ref{fig:pl hists}, distinguished by systems with and without gas giants. Uncertainties on all histogram bins are calculated in a Monte Carlo fashion, where each histogram distribution is computed 1000 times, drawing new masses and radii for each planet from those measured distributions and propagating errors through. 

Starting with the most straightforward characterization of bulk composition, planet densities are given by $\rho = \frac{M_{\mathrm{planet}}}{R_{\mathrm{planet}}^3}$ in units of $\rho_{\oplus}$. From the top panel of Figure~\ref{fig:pl hists}, we see that density distributions of small planets in systems with and without gas giants are broadly similar. There are some hints of differences, including a slight shift in the peak towards higher bulk densities for small planets with gas giants, but a more extended tail towards higher densities for small planets {\it without} gas giants.

\begin{figure*}
    \centering
    \includegraphics[width=0.48\linewidth]{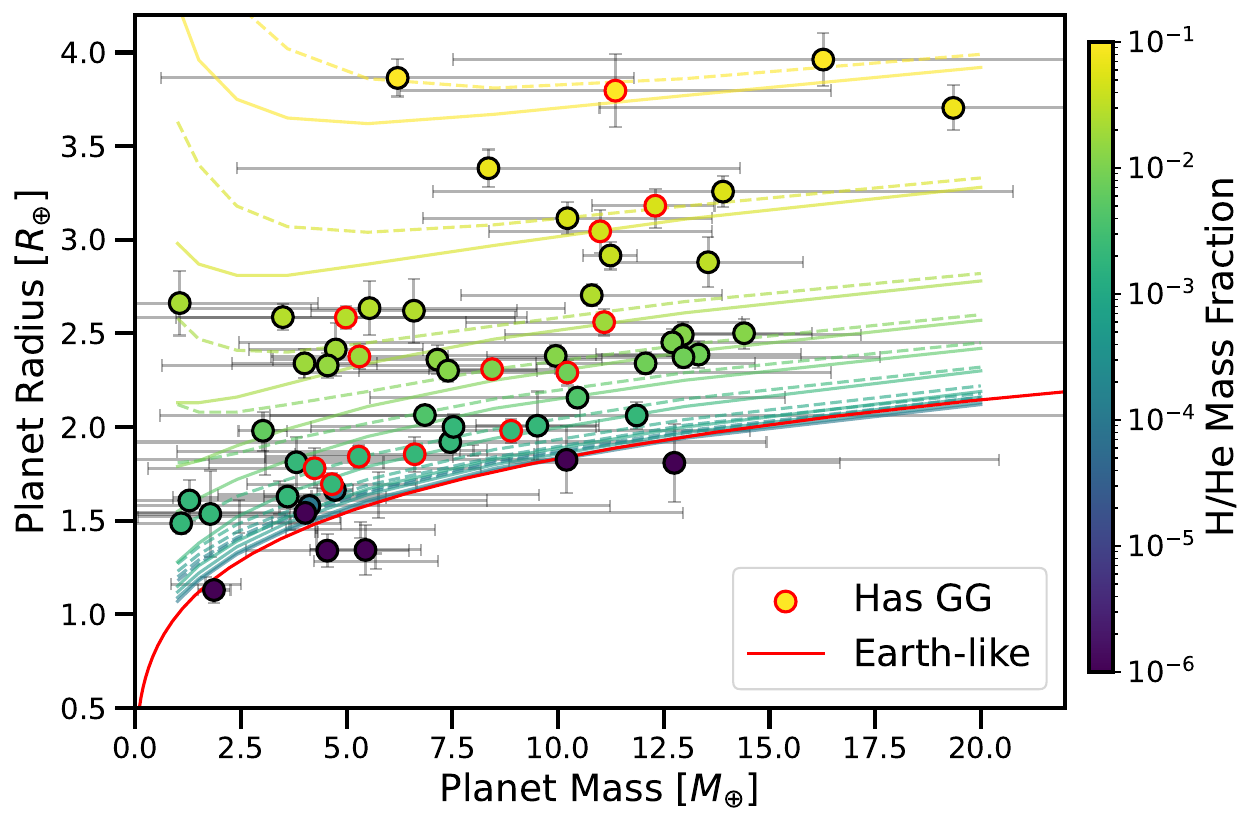}
    \includegraphics[width=0.48\linewidth]{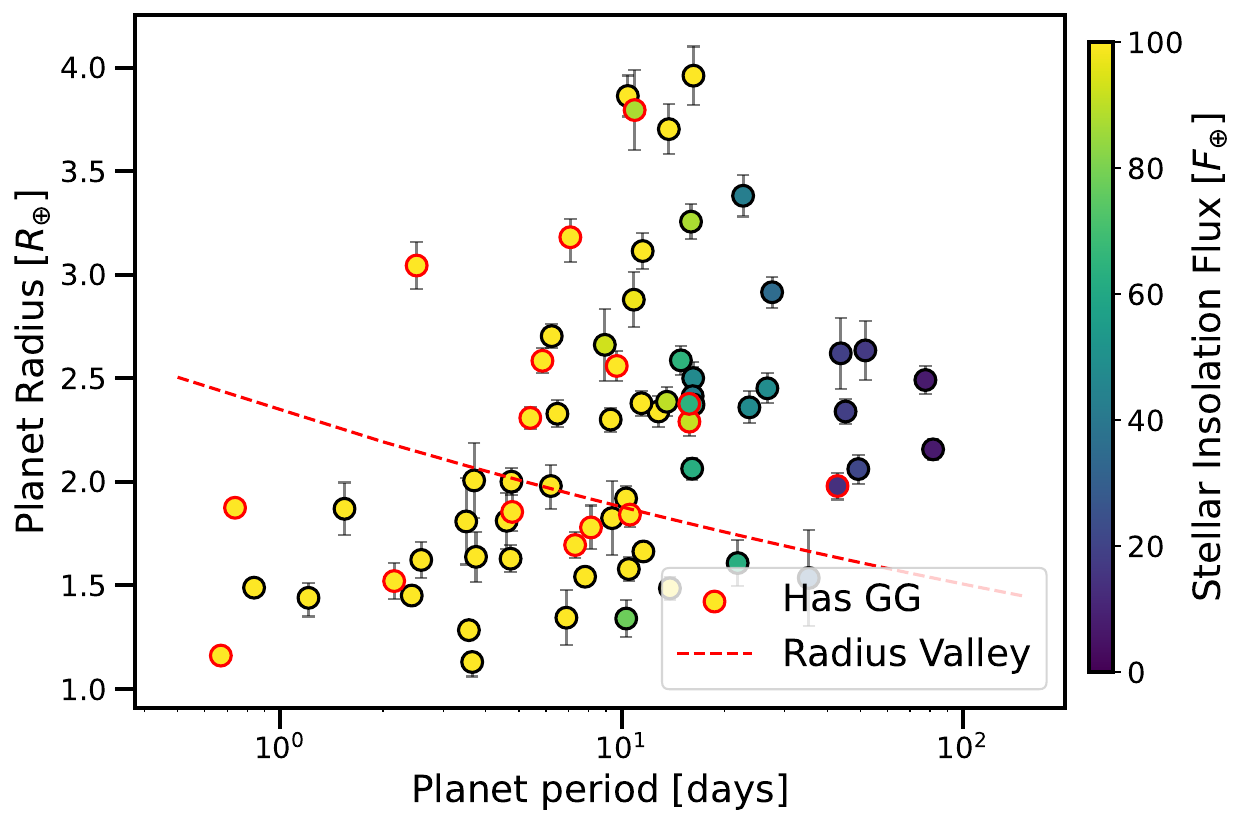}
    
    \caption{\textbf{Left:} Small planet radius vs. mass for our sample, where small planets in systems with outer gas giants are outlined in red. Interpolated H/He envelope fractions from \cite{LopezED2014} are overplotted, with solid curves corresponding to $F_{\mathrm{planet}}= 0.1 F_{\oplus}$, and dashed lines to planets with $10 F_{\oplus}$ assuming an age of 1 Gyr. The solid red line shows an Earth-like planet composition \citep{ZengL2019}. \textbf{Right:} Small planet radius vs. orbital period colored by calculated stellar insolation fluxes. The red dashed line defines the orbital period-dependent radius valley outlined in \cite{HoCSK2023}. 
    }
    \label{fig:rad mass}
\end{figure*}

Next we calculate planet EMF and core mass to disentangle the separate contributions of the atmosphere and the core to the measured mass and radius. We rely on models from \cite{LopezED2014}, which compute radii for low-mass (1--20 M$_{\oplus}$) planets assuming a rocky core and hydrogen-helium envelope. To determine EMFs for our sample, we interpolate these model grids that link a planet radius to a corresponding planet mass, EMF, irradiation, and system age. 

Given the coarse grid of ages considered in \cite{LopezED2014} (100 Myr, 1 Gyr, 10 Gyr), we conservatively assume system ages of 1 Gyr for our sample (although we note selecting 10 Gyr instead does not significantly impact our findings). We calculate insolation fluxes for each planet using published stellar luminosities and semi-major axes. 

For every planet, we select the nearest planet fluxes from Table 2 in \cite{LopezED2014} ($0.1, 10, 1000$ $\text{F}_{\oplus}$). We then construct a linear two-dimensional interpolator for radius $R$ as a function of total mass $M$ and envelope mass fraction, $R(M,\mathrm{EMF})$, using the masses, radii, and EMFs's in their same Table 2. The best EMF is determined by selecting the EMF with the lowest difference between the predicted and observed planet radii. If a planet's radius falls below the mass-radius relation tabulated by \citet{LopezED2014}, we assume that they are rocky and set their EMF to $10^{-6}$, matching that of the Earth. In Figure~\ref{fig:pl hists}, the middle panel shows the distribution of EMF's for small planets in systems with (blue) and without (red) gas giants, with the zone below EMF of 3$\times 10^{-5}$ greyed out as it falls below the available parameter space of \citet{LopezED2014}.
By eye, we see similar distributions of EMFs between systems with and without gas giants. 

Finally, we determine the core masses of each planet by multiplying the planet's core mass fraction (CMF = 1 -- EMF) by its measured mass. These values are shown in the bottom panel of Figure~\ref{fig:pl hists}, which display  comparable core masses of planets in systems with and without gas giants.

Before quantifying occurrence rates, we additionally look for broad trends connected to the presence/absence of an outer GG in 2D mass/radius and radius/period parameter space. The left plot of Figure~\ref{fig:rad mass} displays our sample of planet masses and radii colored by their derived EMFs. The planets with an outer GG (outlined in red) appear evenly distributed in planet mass and radius. The smooth and dashed curves represent the planet masses, radii, and EMFs provided in Table 2 of \cite{LopezED2014} for a 0.1 F$_{\oplus}$, and 10 F$_{\oplus}$ planet respectively. Our derived planet EMFs lie near or on top of these curves indicating that our interpolation has worked accordingly. The red line is taken from \cite{ZengL2019} and represents an Earth-like planet ($32.5\%$ Fe, $67.5\%$ MgSi$\text{O}_{3}$) for reference. 

In the right panel of Figure~\ref{fig:rad mass}, we display our small planets' radii as a function of their orbital periods colored by their derived insolation fluxes. As expected, planets with longer periods have lower insolation fluxes. The radius valley derived in \cite{HoCSK2023} is plotted as the red dashed line. By eye there does not appear to be a preference for small planets with an outer GG to lie above or below this boundary. We revisit this question in Sections \ref{sec: no diff valley} and \ref{sec:discussion}.

\begin{figure*}
    \centering
    \includegraphics[width=0.32\linewidth]{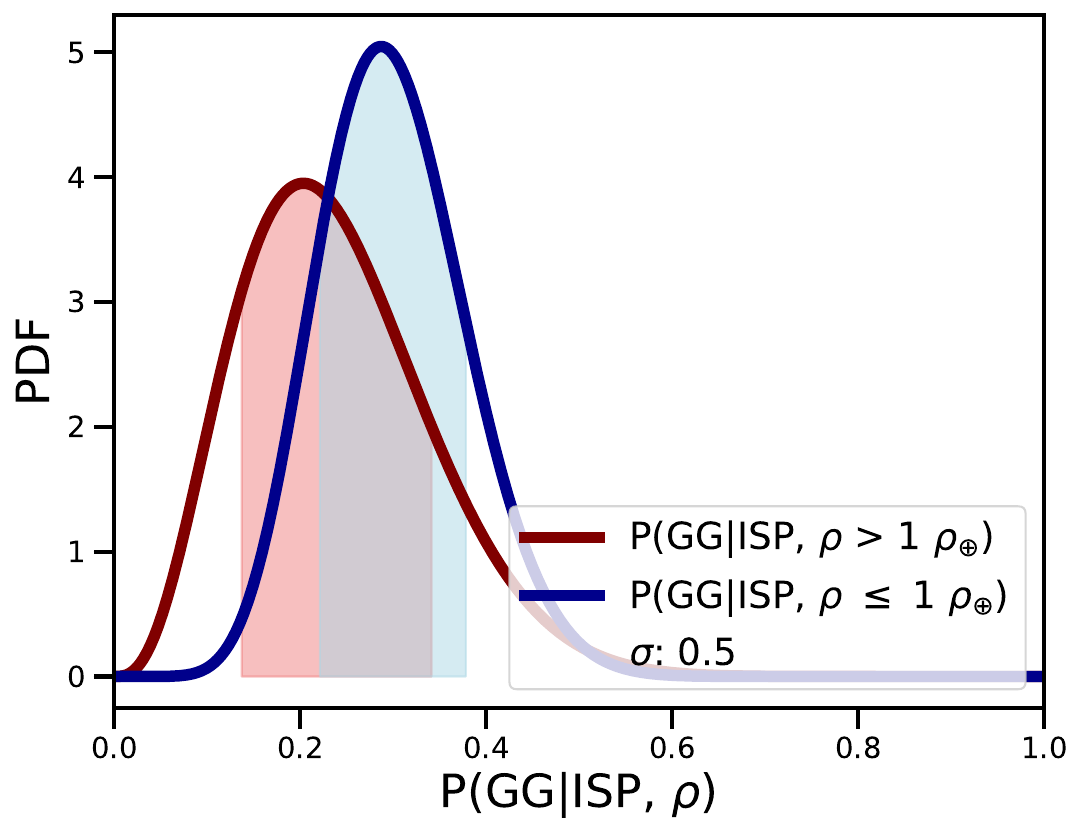}
    \includegraphics[width=0.32\linewidth]{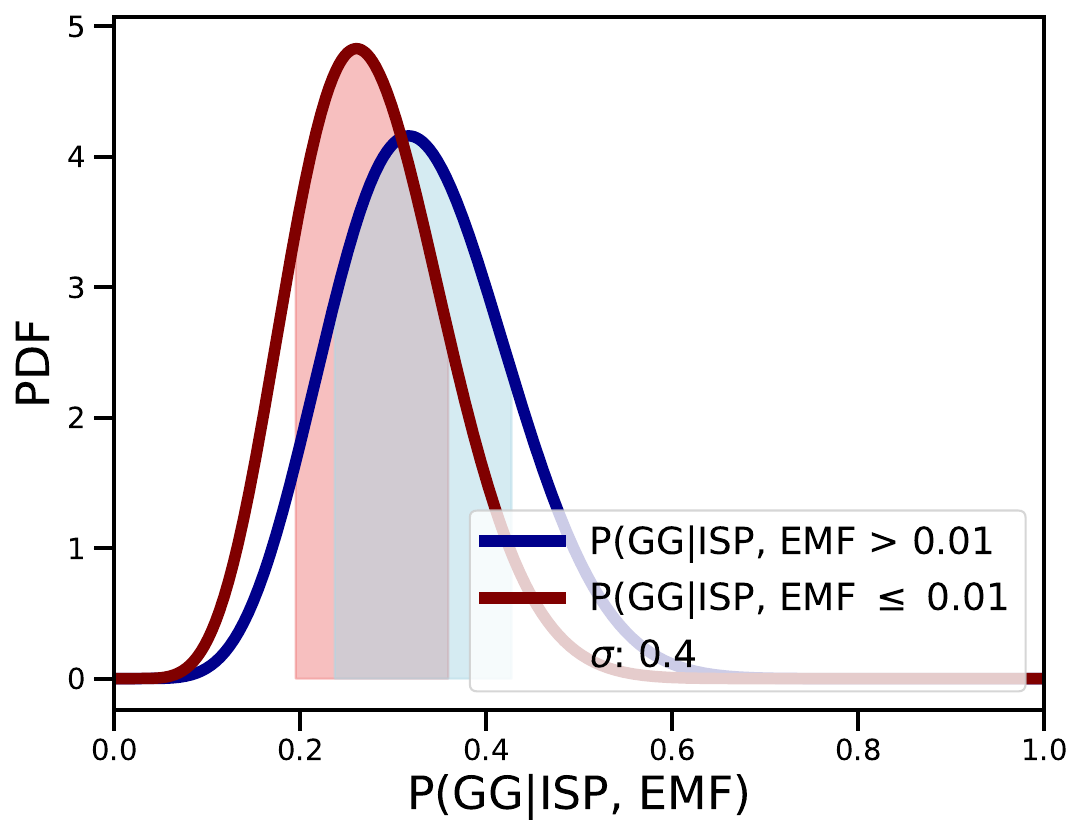}
    \includegraphics[width=0.32\linewidth]{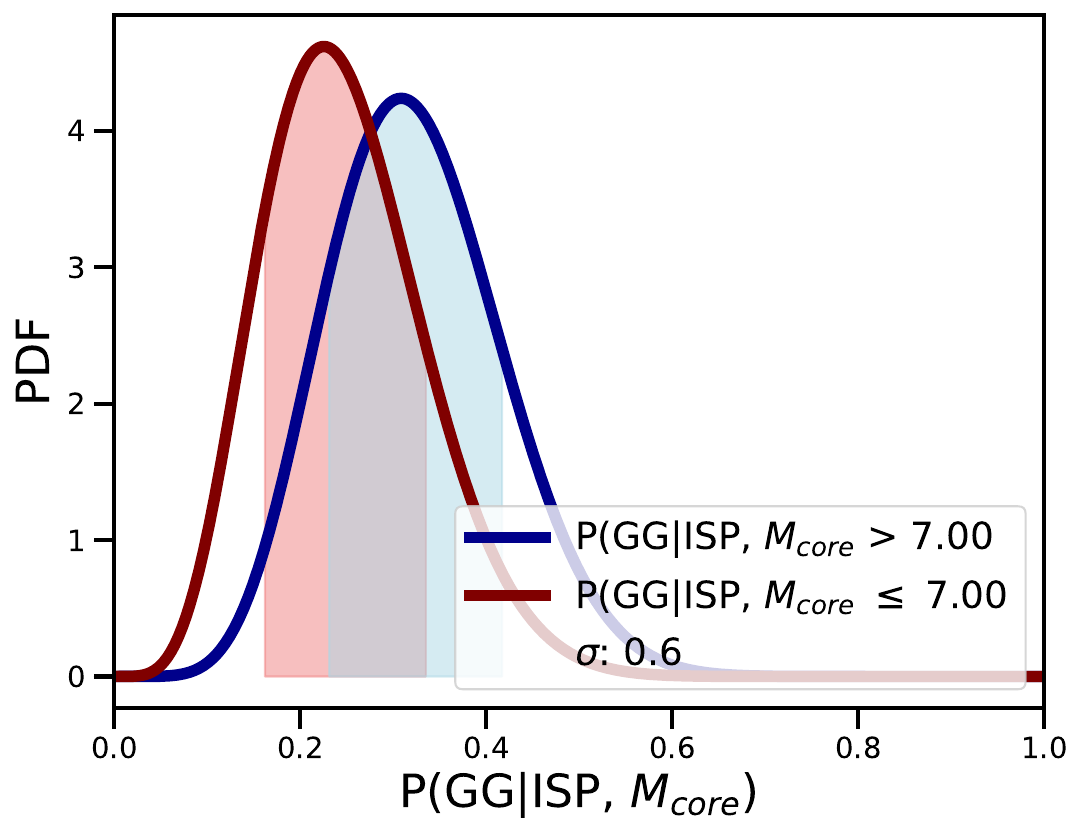}
    
    \includegraphics[width=0.32\linewidth]{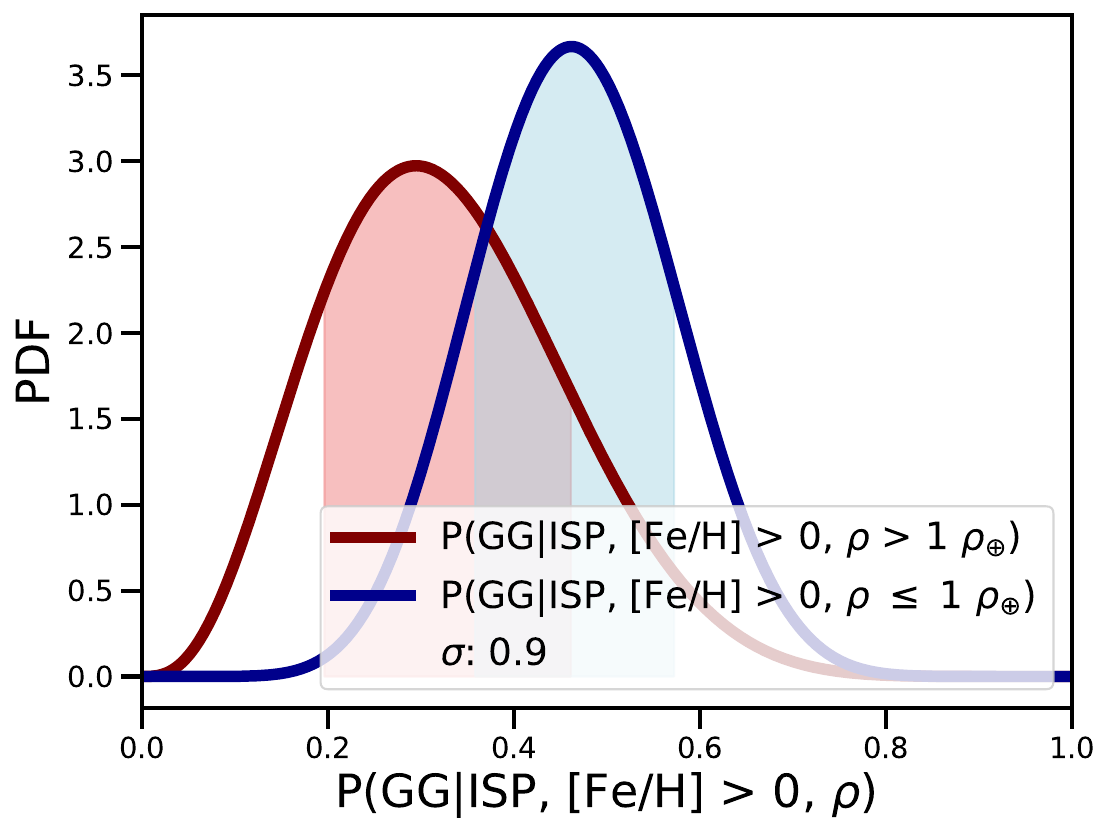}
    \includegraphics[width=0.32\linewidth]{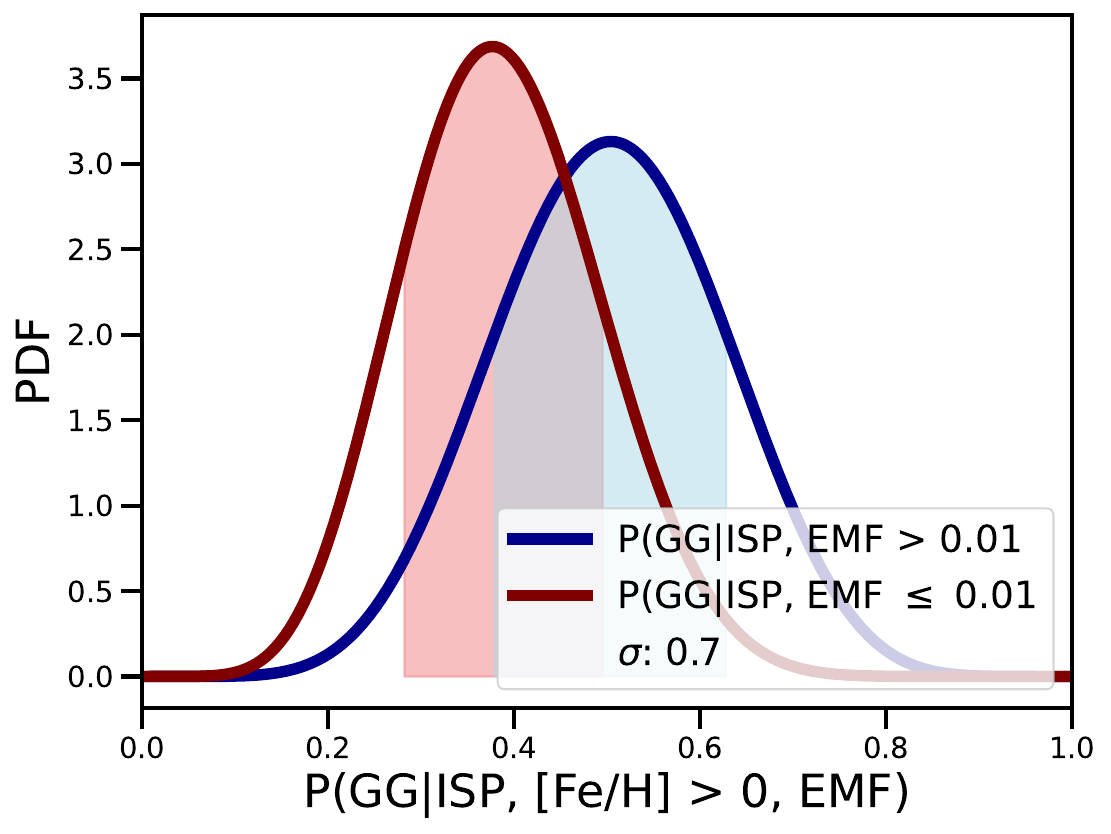}
    \includegraphics[width=0.32\linewidth]{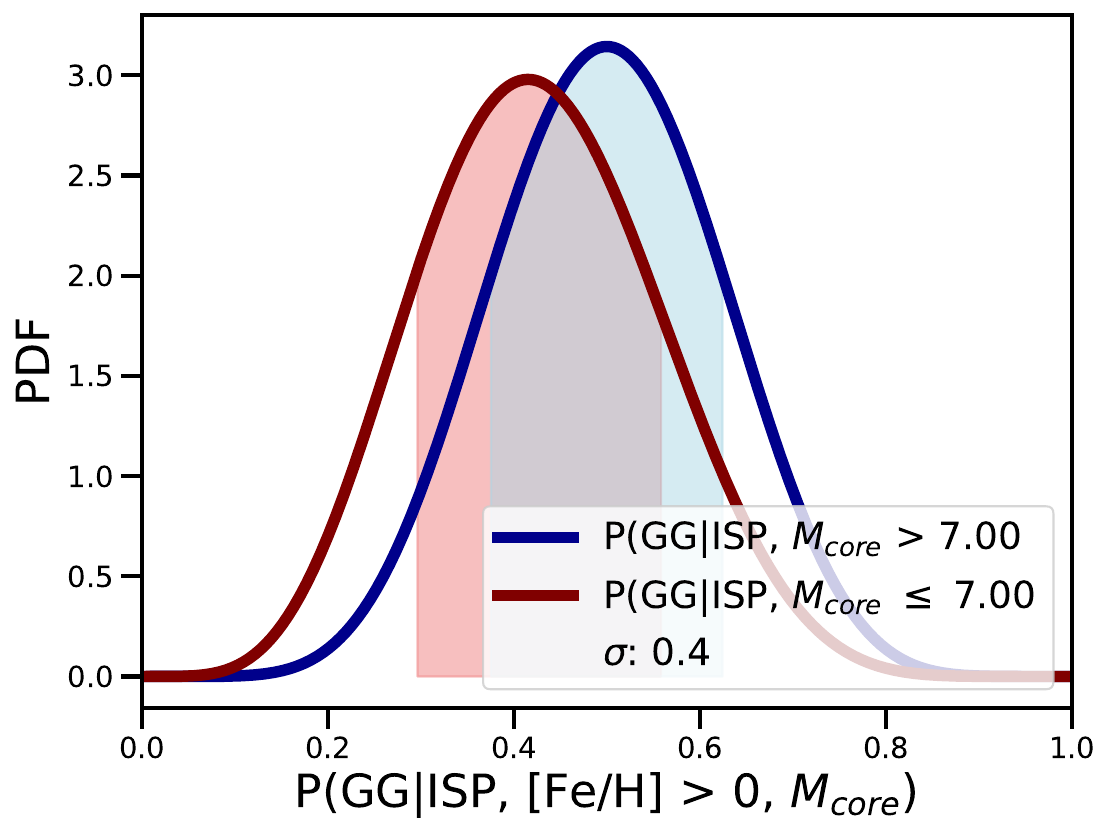}

    \includegraphics[width=0.32\linewidth]{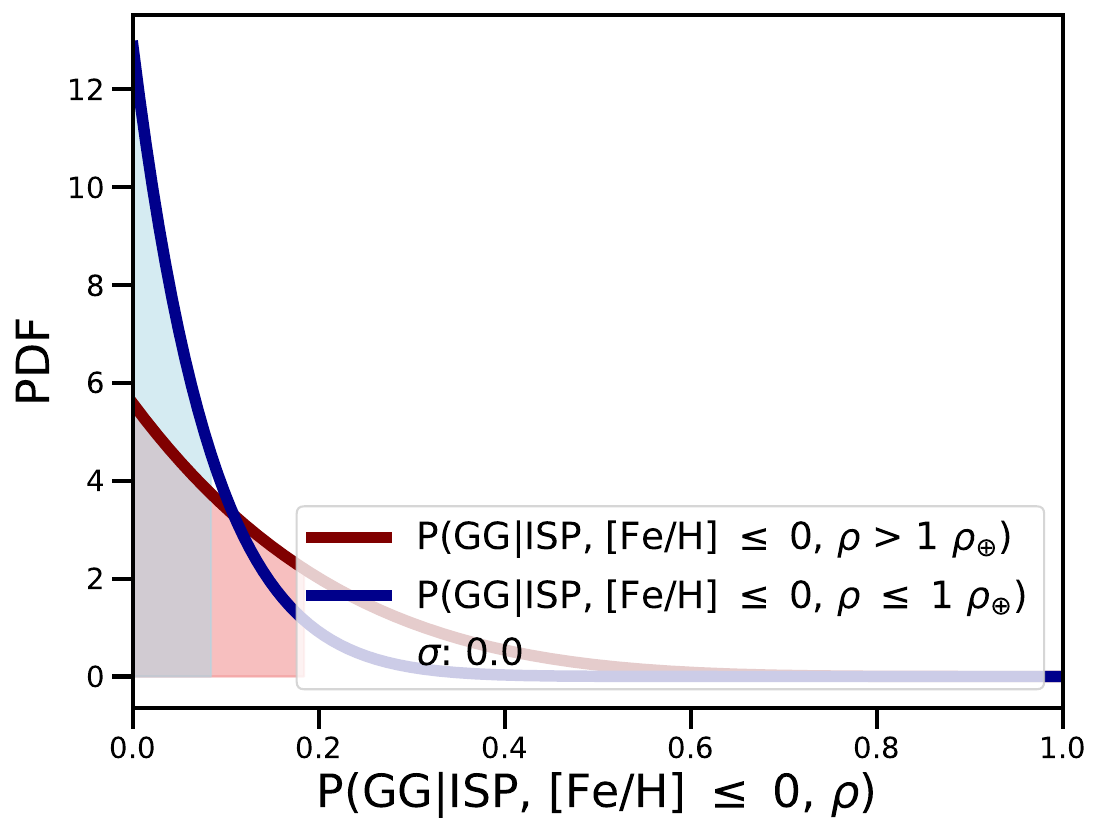}
    \includegraphics[width=0.32\linewidth]{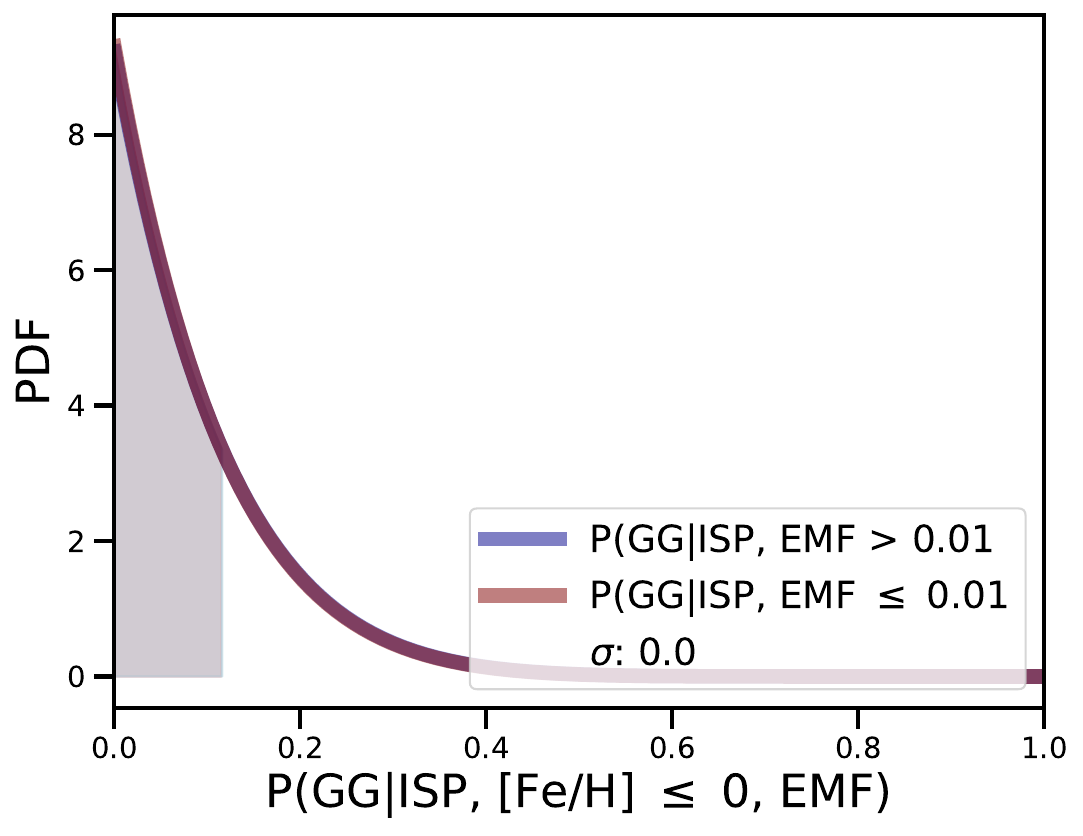}
    \includegraphics[width=0.32\linewidth]{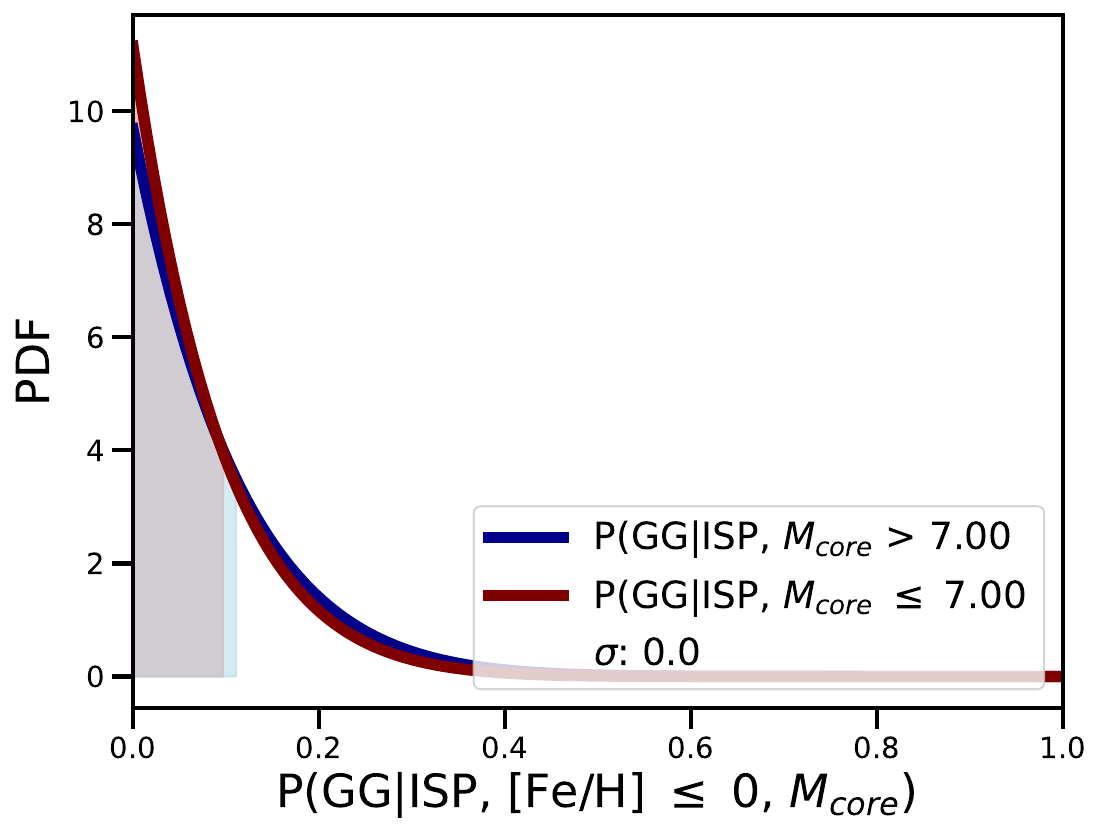}
    \caption{Conditional occurrence rates P(GG$|$ISP) split by density (\textbf{left column}), EMF (\textbf{middle column}), and core mass (\textbf{right column}) for the entire sample (\textbf{top row}), metal-rich systems (\textbf{middle row}), and metal-poor systems (\textbf{bottom row}). The shaded region indicates the 68\% confidence interval for each distribution. Densities were divided at 1$\rho_{\oplus}$, EMF at a value of 1$\%$, and core mass at 7 M$_{\oplus}$.}
    \label{fig:probability distributions}
\end{figure*}

We now have all the ingredients to quantify how the occurrence rate of gas giants in inner small planet systems P(GG$|$ISP) varies as a function of small planet density, EMF, and core mass. Following methodologies outlined in Section~\ref{sec:sanity check}, we calculate the conditional occurrence rates using: 1) our entire sample; 2) just the metal-rich subsample; and lastly 3) the metal-poor sample. We show all probability distributions in Figure~\ref{fig:probability distributions}, where the top row shows our entire sample of 43 systems, the middle row contains just the metal-rich [Fe/H]$>$0 systems, and the bottom row represents the metal-poor [Fe/H]$\leq$0 systems. Corresponding occurrence rates are tabulated in Table \ref{table: occ rates}. 

\begin{deluxetable*}{lcccccc} \label{table: occ rates}
\tablecaption{Occurrence Rates of Outer Gas Giants}
\scriptsize
\tablehead{
\colhead{Conditional Probability} & \colhead{$\rho$ $>$ 1$\rho_{\oplus}$} & \colhead{$\rho$ $\leq$ 1$\rho_{\oplus}$} & \colhead{EMF $>$ 1$\%$} & \colhead{EMF $\leq$ 1$\%$} & \colhead{M$_{\rm core}$ $>$ 7M$_{\oplus}$} & \colhead{M$_{\rm core}$ $\leq$ 7M$_{\oplus}$}
}
\startdata
P(GG$|$ISP) & 20.3$^{+13.8}_{-6.6}$$\%$&28.7$^{+9.1}_{-8.0}$$\%$& 31.7$^{+11.0}_{-8.0}$$\%$& 26.0$^{+9.9}_{-6.5}$$\%$ & 30.8$^{+10.9}_{-6.5}$$\%$ & 22.5$^{+11.0}_{-6.3}$$\%$\\[1ex]
P(GG$|$ISP, [Fe/H]$>$0) & 29.5$^{+16.6}_{-9.9}$$\%$ & 46.2$^{+11.0}_{-10.4}$$\%$ & 50.4$^{+12.4}_{-12.5}$$\%$ & 37.6$^{+11.9}_{-9.4}$$\%$ & 50.0$^{+12.4}_{-12.5}$$\%$ & 41.5$^{+14.3}_{-11.9}$$\%$\\[1ex]
P(GG$|$ISP, [Fe/H]$\leq$0) & $<$ 18.4$\%$ & $<$ 8.4$\%$ & $<$ 11.6$\%$& $<$ 11.5$\%$ & $<$ 11.1$\%$ & $<$ 9.7$\%$\\
\enddata
\end{deluxetable*}

Each column of Figure~\ref{fig:probability distributions} illustrates how the frequency of gas giants in small planet systems depends on planet density (left column), EMF (middle column), and core mass (right column). Since we aim to determine whether gas giants are more or less common in systems \textit{as a function of} our three bulk composition properties, we take the simplest option and define a dividing line for ``high'' and ``low'' planet density (1$\rho_\oplus$), EMF (1\%), and core mass (7$M_\oplus$). 

Our choice of the value of EMF and core mass at which we make the division splits the sample into approximately equal numbers of systems on either side. For planet density we simply choose a split at the density of the Earth, which yields a slightly larger low-density sample than high-density sample, but we note that our conclusions remain consistent regardless.
 
Since we are calculating the occurrence of gas giants in systems with inner small planets falling on either side of the aforementioned dividing lines, one can imagine a scenario in which there is an outer gas giant in a system with two inner small planets, each of which falls on opposite sides of one of the dividing lines (e.g., one is ``high'' density one ``low'' density). In these cases we count this system twice: in our occurrence rate calculation this hypothetical system would count as both a low-density small planet system with an outer gas giant, as well as a high-density small planet system with an outer gas giant. This double counting preserves the relative comparison of occurrence rates, which is the objective.

When considering our entire sample (top row of Figure~\ref{fig:probability distributions}), for all planet parameters the frequencies of gas giants in systems with small planets above and below the dividing parameter value are consistent to $\sim 0.5\sigma$. Similarly, we see no difference in P(GG$|$ISP) as a function of inner planet parameters in metal-poor systems (bottom row Figure~\ref{fig:probability distributions}), which is essentially driven by the fact that there are no gas giant detections in these systems.

When we move to the metal-rich systems, we first note that for each panel the probability distributions become broader than the top row due to the smaller sample size (27 vs.~43 systems). Upon further inspection, even with these less precise constraints we see the significance of the difference between the occurrence of gas giants in small planet systems with high/low densities and high/low EMFs \textit{increases}. Metal-rich systems with less dense small planets are more likely to have an outer gas giant to 0.9$\sigma$ significance, compared to 0.5$\sigma$ for the entire sample. Similarly, metal-rich systems with small planets that have higher EMFs are more likely to host an outer gas giant by 0.7$\sigma$, in comparison to 0.4$\sigma$ for the entire sample. By contrast, no significant enhancement in P(GG$|$ISP) is observed for higher core masses in metal-rich systems compared to the all metallicity sample; in fact, the significance decreases.
Nevertheless, in all cases the significance of enhancement in P(GG$|$ISP) remains below 1$\sigma$. A more definitive test would require more mass measurements of the short-period small planets. With the current sample, we contend with a potential hint rather than a firm conclusion.

\section{Sample division by radius valley and total mass}
\label{sec: no diff valley}

\subsection{No significant difference in GG occurrence}

Another way to divide a sample of inner small planets into high vs.~low density is make use of the radius valley, which is a clear division between the smaller rocky super-Earths (SE; below the valley) and the larger volatile-laden sub-Neptunes (SN; above the valley) \citep{FultonBJ2017,FultonBJ2018,vaneylen2018}. While a less precise method of probing planet composition, division across a radius valley has the advantage of increasing the base sample size as we no longer require the inner planets to have both radius and mass measurements.

We start with the 184 SE/SN systems in \citet{BryanML2024}. Here we account for several more recent updates to the sample, including HD 26965 that is confirmed to be false positive, and WASP-132 that has a recently published outer gas giant. With these minor updates, we have 56 metal-rich systems with transiting planets 1--4 R$_\oplus$, 17 of which have outer gas giants, and 55 metal-poor systems with transiting planets 1--4 R$_\oplus$, 2 of which have outer gas giants. 

We adopt equation 11 in \citet{HoCSK2023} as the radius valley to divide our sample of systems with measured small planet radii:

\begin{equation}
\begin{split}
    \log_{10}\left(\frac{R_p}{R_\oplus}\right) = &-0.09\log_{10}\left(\frac{P}{\rm days}\right) \\
    &+ 0.21\log_{10}\left(\frac{\rm M_\star}{\rm M_\odot}\right) + 0.35.
\end{split}
\end{equation}
In the case where a single system harbors planets that fall both below and above the radius valley, we double count this system, placing it in both the SE sample and the SN sample when calculating respective occurrence rates of outer gas giants. This double counting preserves the relative comparison between SE and SN systems, an implication that we discuss further in Section \ref{sec:caution}. 

\begin{figure*}
    \centering
    \includegraphics[width=0.32\linewidth]{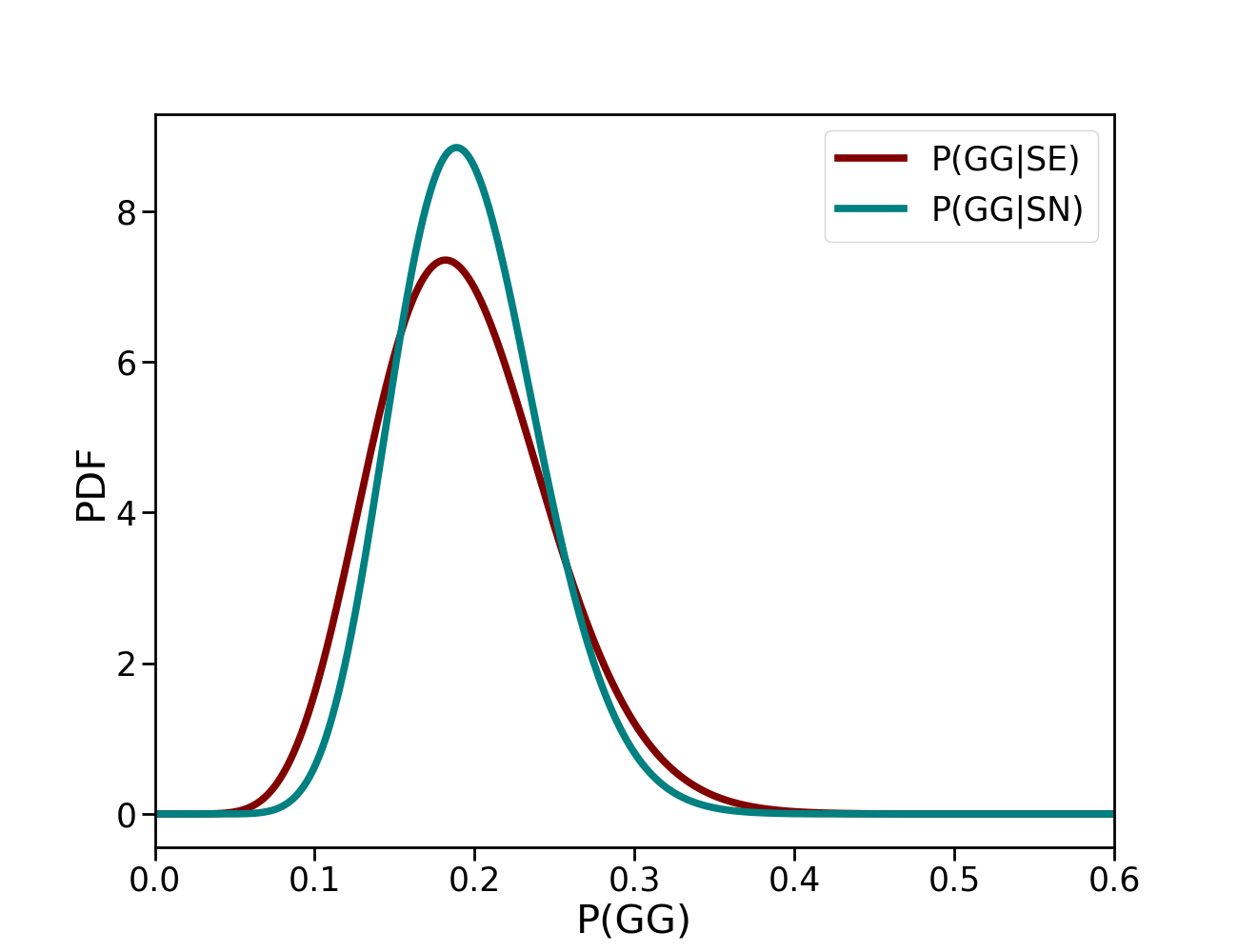}
    \includegraphics[width=0.32\linewidth]{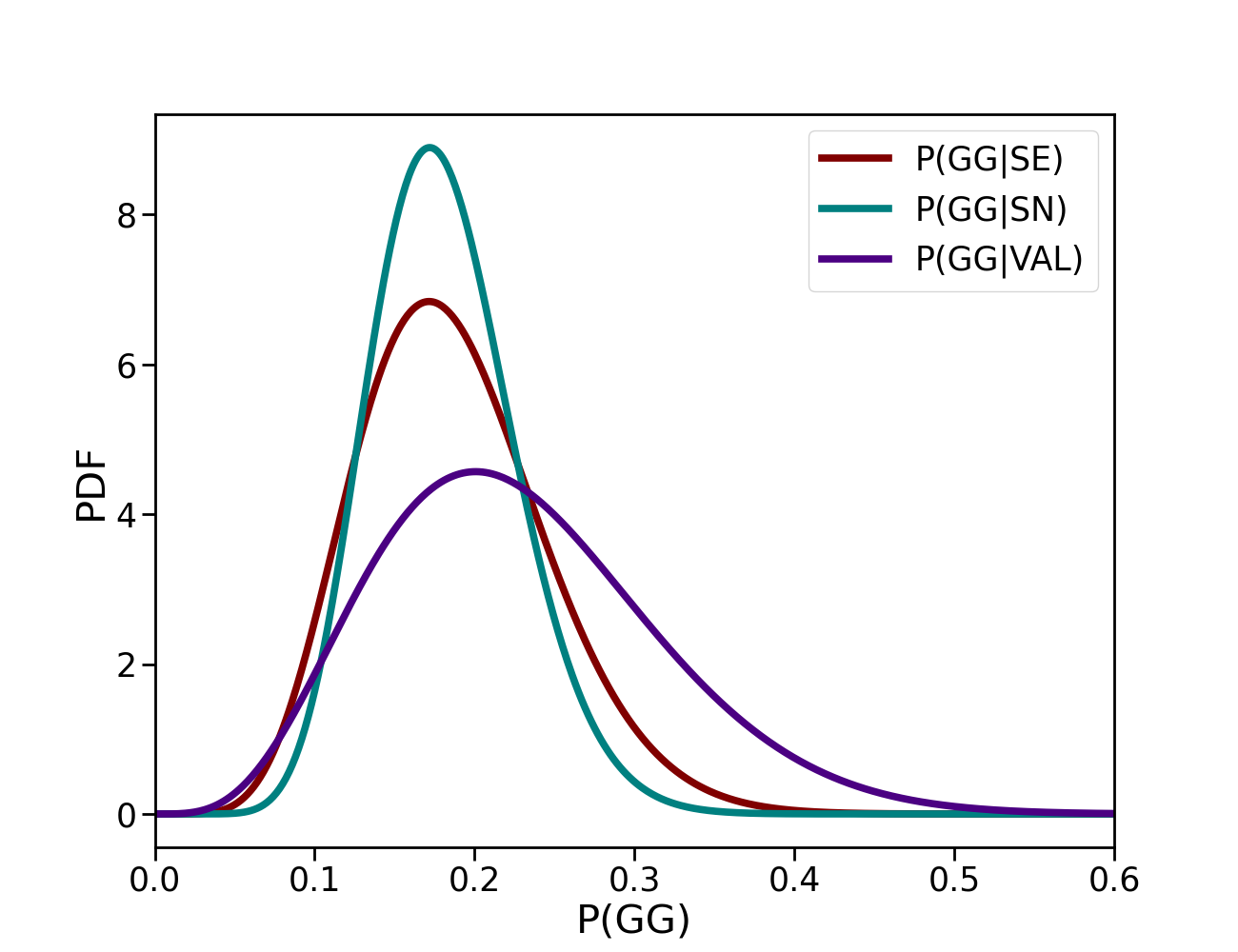}
    \includegraphics[width=0.32\linewidth]{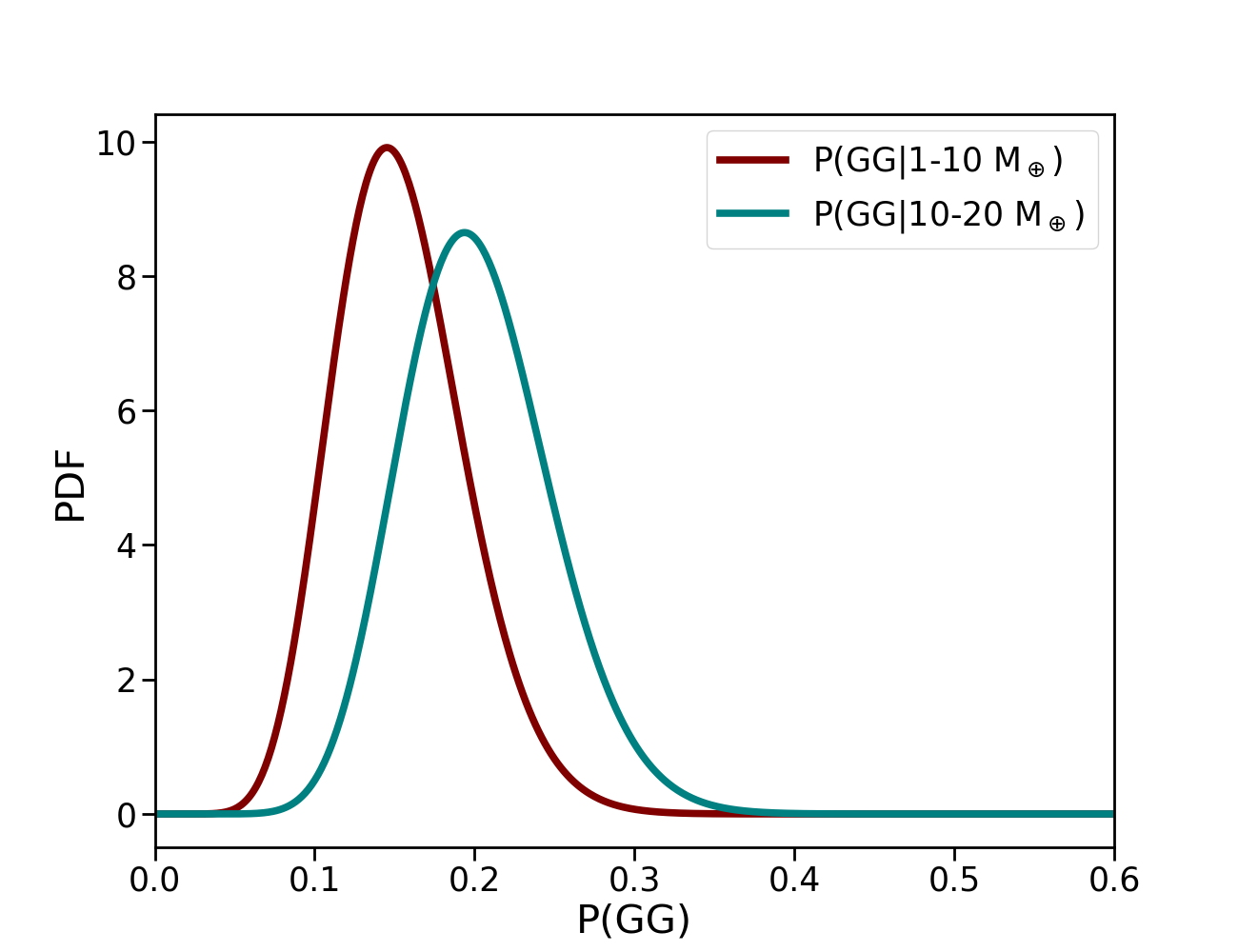}
    
    \includegraphics[width=0.32\linewidth]{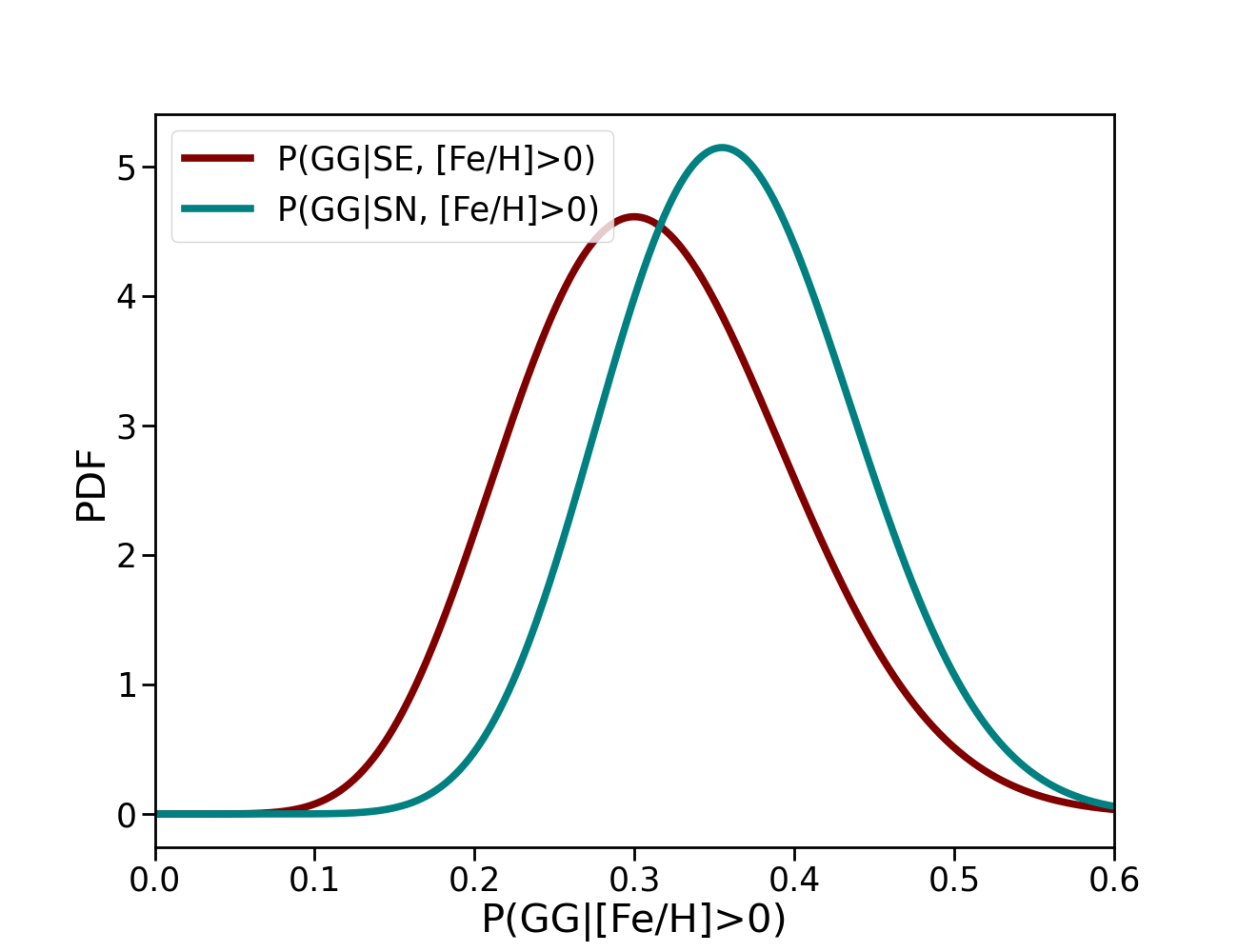}
    \includegraphics[width=0.32\linewidth]{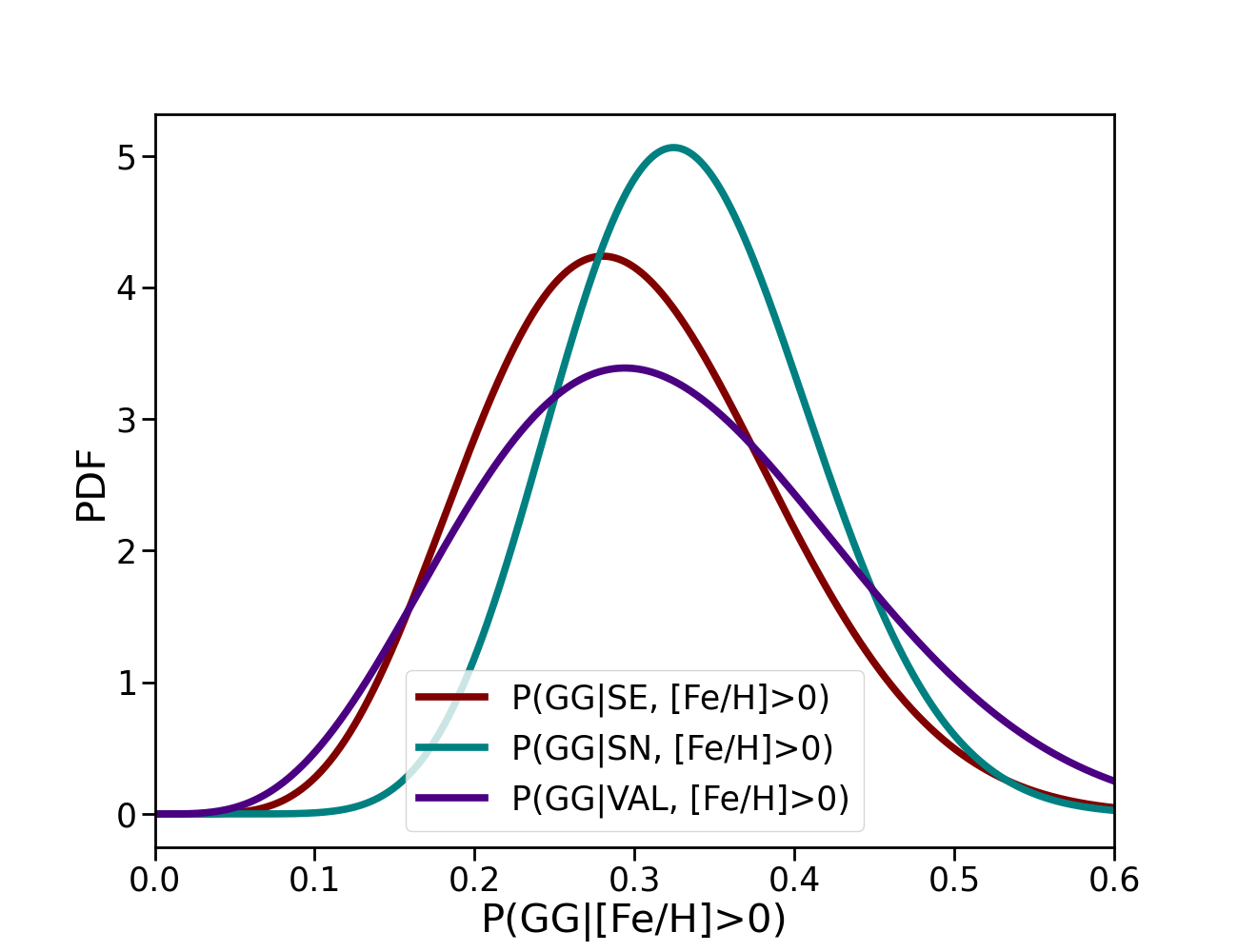}
    \includegraphics[width=0.32\linewidth]{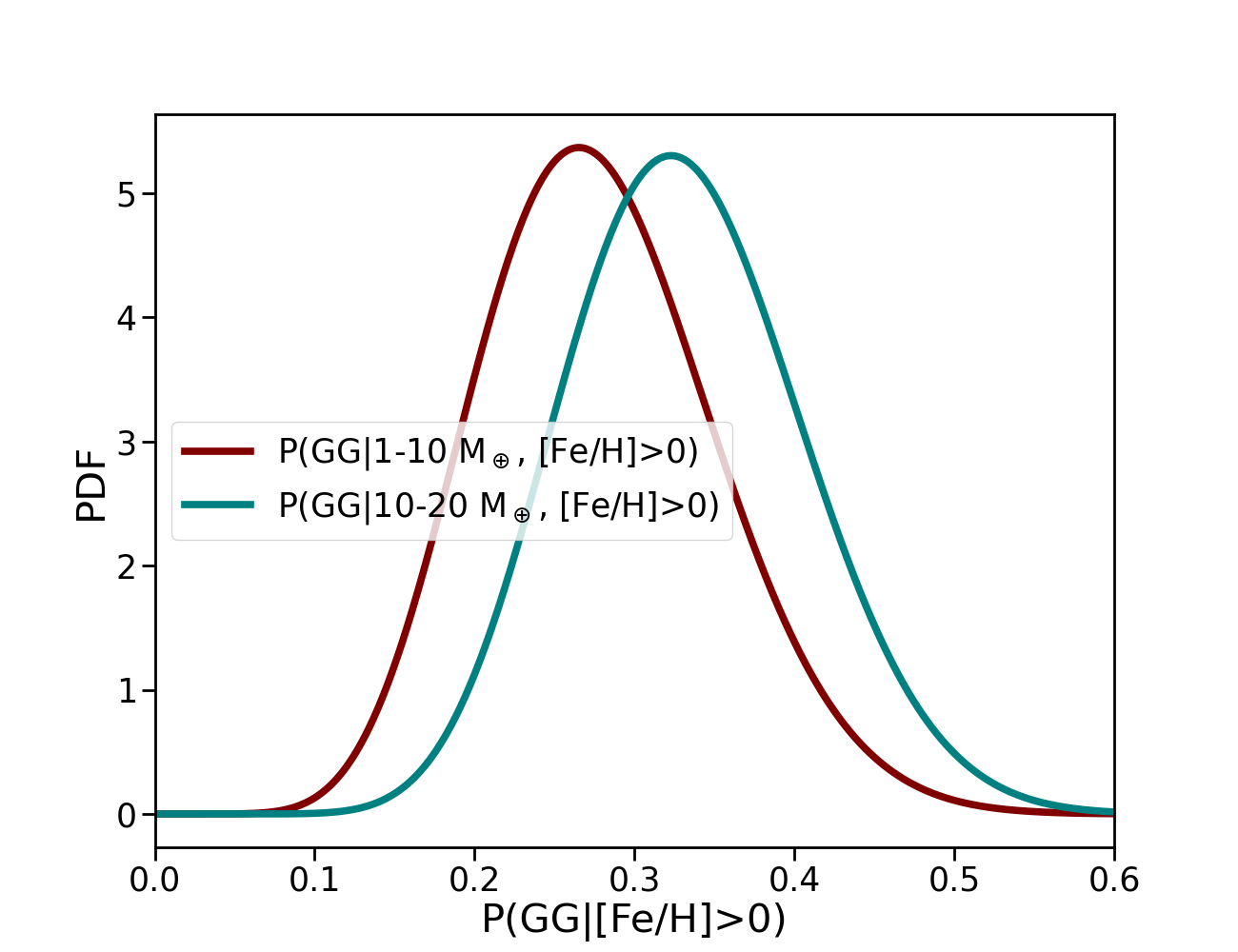}
    
    \includegraphics[width=0.32\linewidth]{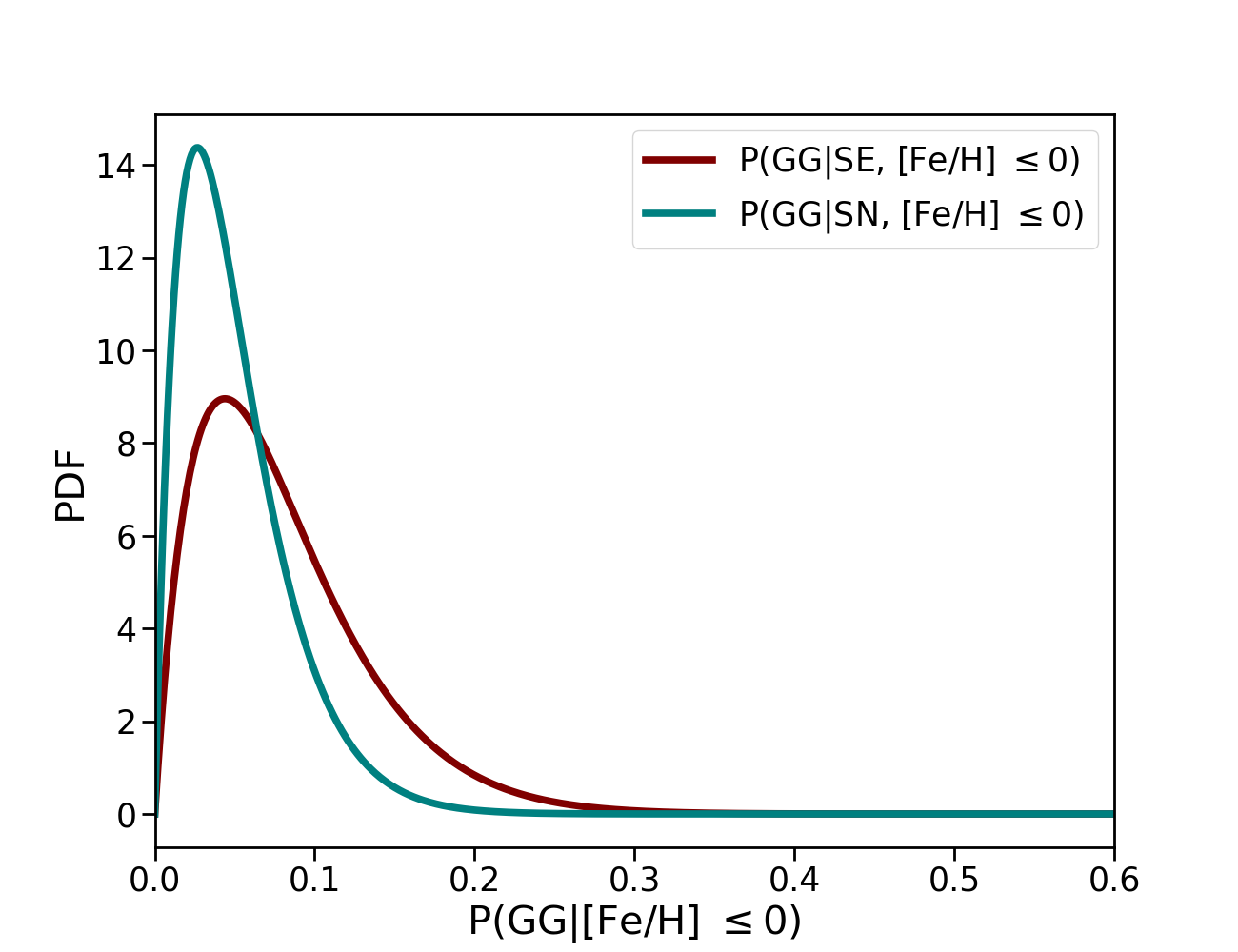}
    \includegraphics[width=0.32\linewidth]{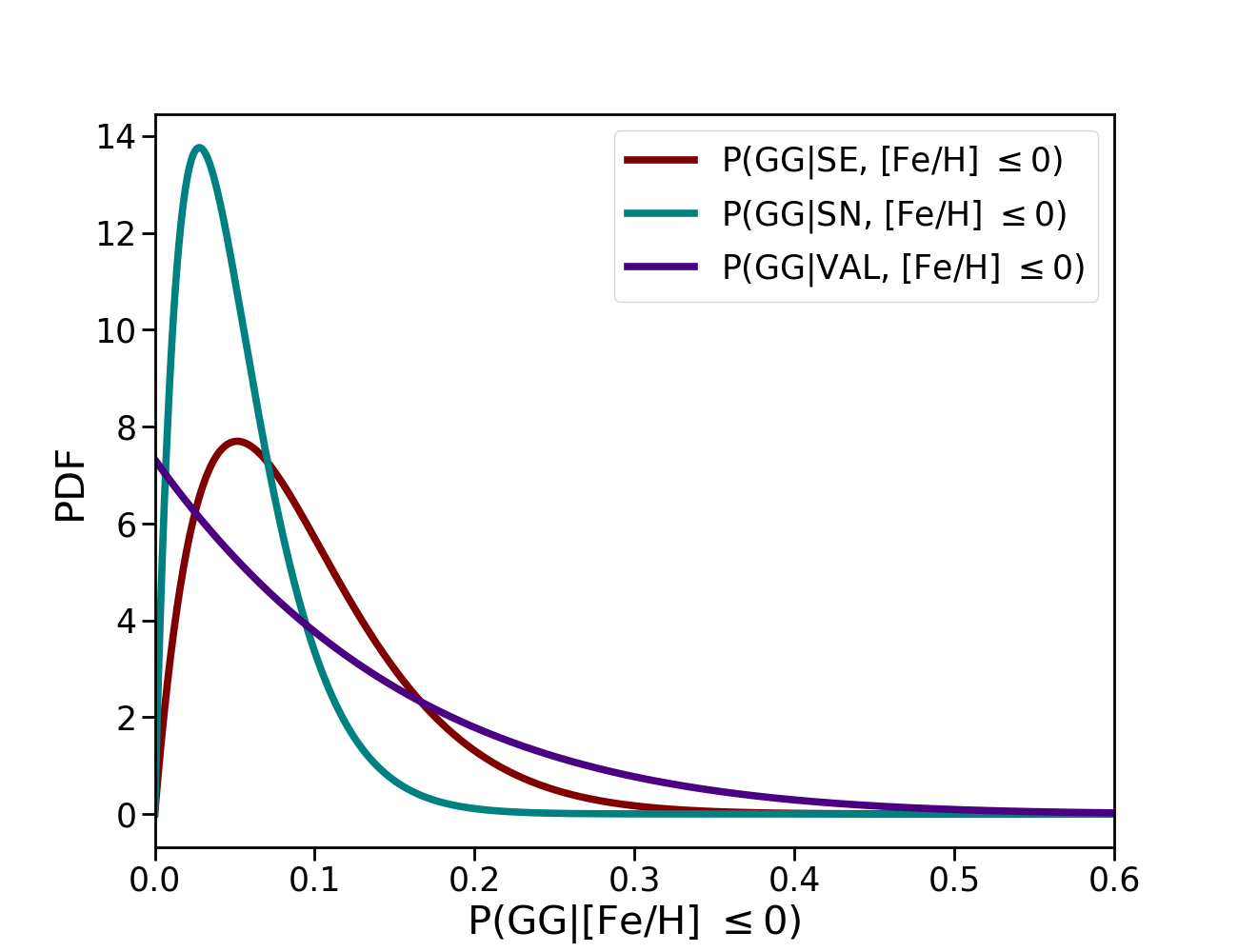}
    \includegraphics[width=0.32\linewidth]{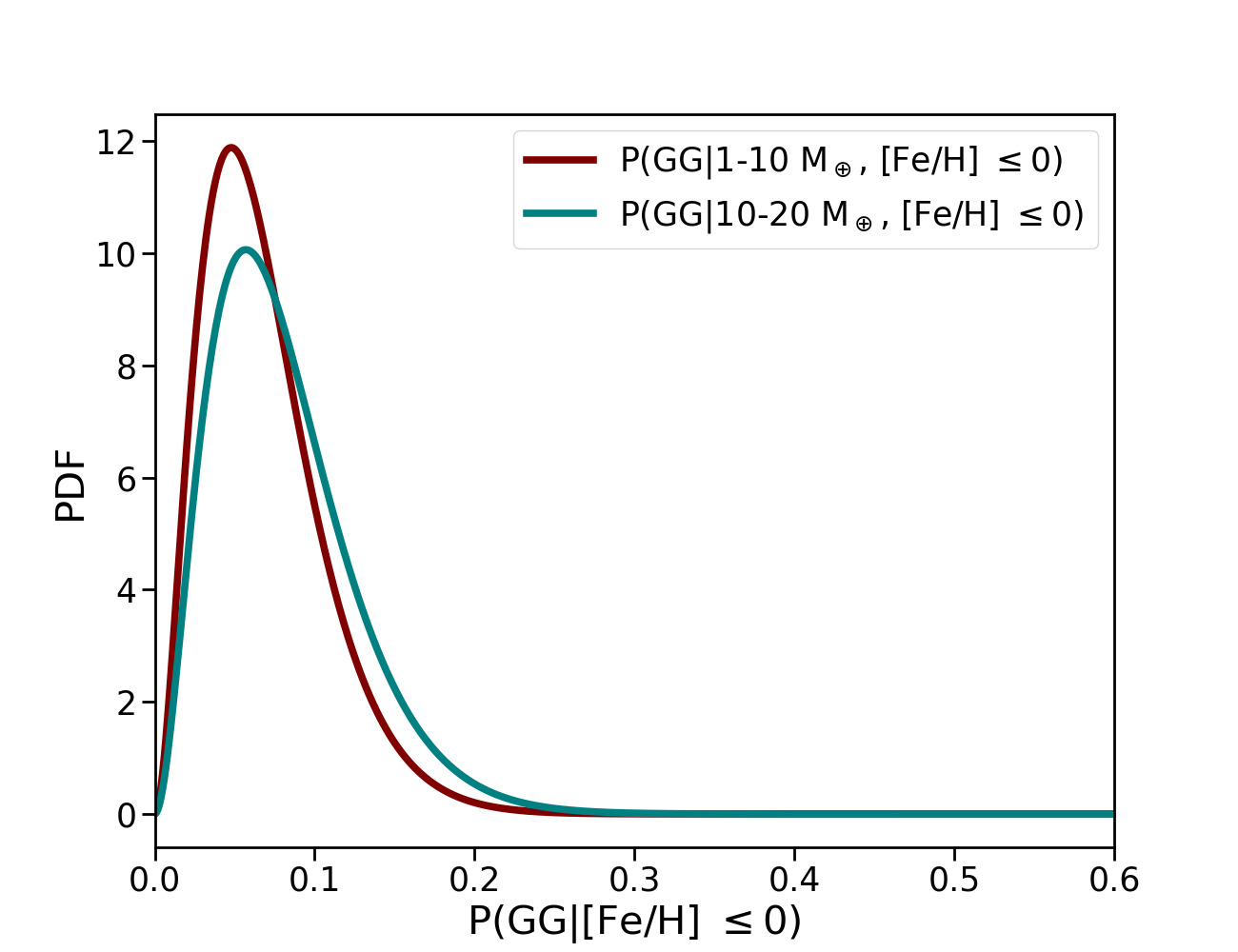}
    
    \caption{\textbf{Left column}: Conditional occurrence rates of outer gas giants in systems with inner super-Earths (SE) or sub-Neptunes (SN). SE/SN are categorized using the \citet{HoCSK2023} equation 11 whereby SE fall below the valley and SN are above. Top, middle, and bottom panels show this frequency for all, high, and low metallicity systems respectively. \textbf{Middle column}: Separating out planets that fall in the radius valley, width defined by dashed line boundaries in Figure 2 of \citet{HoCSK2023}. Super-Earths (SE) are planets that fall below this lower boundary, sub-Neptunes (SN) fall above the upper boundary, and valley planets (VAL) fall between the boundaries, i.e. in the radius valley. \textbf{Right column}: Conditional occurrence rates of outer gas giants in systems with inner planets 1--10 M$_\oplus$ or planets 10--20 M$_\oplus$. Top, middle, and bottom panels show this frequency for all, high, and low metallicity systems respectively.}
    \label{fig:rad vally and mass cuts}
\end{figure*}

After computing individual system sensitivities to distant giants per Section \ref{sec:injection recovery}, we calculate occurrence rates of outer gas giants in systems hosting super-Earths and systems with sub-Neptunes following Section \ref{sec:sanity check}. The left column of Figure \ref{fig:rad vally and mass cuts} shows how P(GG$|$SE) and P(GG$|$SN) compare to each other given all system metallicities, just the metal-rich systems, and only the metal-poor ones, and Table \ref{table: occ rates r and m} lists the corresponding values. In all three cases we do not find statistically significant differences between the frequencies of outer gas giants in systems with SE vs SN. However, while the combined metallicity sample shows consistency to $<$0.1$\sigma$, the metal-rich one has P(GG$|$SN) higher than P(GG$|$SE) by 0.5$\sigma$, while the metal-poor one has P(GG$|$SE) higher than P(GG$|$SN) by 0.3$\sigma$. Once again the increase in significance despite the decrease in sample size hints at a potential correlation between sub-Neptunes, outer gas giants, and metallicity.

\begin{deluxetable}{lccc} 
\label{table: occ rates r and m}
\tablecaption{GG occurrence given inner planet radius/mass}
\scriptsize
\tablehead{
\colhead{Conditional Probability} & \colhead{All [Fe/H]} & \colhead{[Fe/H]$>$0} & \colhead{[Fe/H]$\leq$0}
}
\startdata
P(GG$|$SE)\tablenotemark{1} & 18.2$^{+6.7}_{-4.2}\%$ & 30.0$^{+10}_{-7.2}\%$ & 4.4$^{+8.8}_{-1.4}\%$\\
P(GG$|$SN)\tablenotemark{1} & 18.9$^{+5.3}_{-3.7}\%$ & 35.5$^{+8.4}_{-6.9}\%$ & 2.7$^{+5.6}_{-0.8}\%$\\
P(GG$|$SE)\tablenotemark{2} & 17.2$^{+7.4}_{-4.3}\%$ & 28.0$^{+11.2}_{-7.5}\%$ & 5.2$^{+10.1}_{-1.6}\%$\\
P(GG$|$SN)\tablenotemark{2} & 17.2$^{+5.4}_{-3.6}\%$ & 32.4$^{+8.8}_{-6.8}\%$ & 2.8$^{+5.9}_{-0.9}\%$\\
P(GG$|$VAL)\tablenotemark{2} & 20.1$^{+11.5}_{-6.1}\%$ & 29.4$^{+14.3}_{-9.0}\%$ & $<$22.1$\%$\\
P(GG$|$1--10 M$_\oplus$) & 14.5$^{+5.0}_{-3.1}\%$ & 26.5$^{+8.7}_{-6.7}\%$ & 4.8$^{+5.7}_{-1.6}\%$\\
P(GG$|$10--20 M$_\oplus$) & 19.4$^{+5.4}_{-3.8}\%$ & 32.3$^{+8.3}_{-6.7}\%$ & 5.6$^{+6.7}_{-1.9}\%$\\
\enddata
\tablenotetext{1}{SN/SE are defined by whether the planet radius falls above (SN) or below (SE) the radius valley as defined in \citet{HoCSK2023}.}
\tablenotetext{2}{SN/SE/VAL are defined by whether the planet radius falls above the upper bound of the radius valley (SN; upper dashed line in Figure 2 of \citet{HoCSK2023}), below the lower bound of the radius valley (SE; lower dashed line in Figure 2 \citet{HoCSK2023}), or between these boundaries (VAL; planets in the valley).}
\end{deluxetable}

Alternatively, the planet's total mass can also be used on its own as an approximate, albeit imperfect, way to divide between high vs.~low density planets. \citet{parc2024} show that the observed planets consistent with rocky composition do not exceed 10 M$_\oplus$ whereas puffier planets can be more massive (and also less massive). Defining inner small planets by mass 1--20 M$_\oplus$, here we have 134 small planet systems with mass measurements to $<$30$\%$ precision, among which 65 are around metal-rich stars and 69 around metal-poor hosts. Of the 22 outer gas giants in these systems, 19 are found in the metal-rich systems and 3 in the metal-poor ones.

We divide this total mass sample in two, considering ``low-mass'' 1--10 M$_\oplus$ and ``high-mass'' 10--20 M$_\oplus$ planets where the masses have been measured with precision $<$30$\%$. Again if a single system contains both a low-mass and high-mass planet, we double count the system in both the 1--10 M$_\oplus$ subsample and the 10--20 M$_\oplus$ subsample. The right column of Figure \ref{fig:rad vally and mass cuts} shows the probability distributions for all, high, and low metallicities comparing occurrences of gas giants in systems with low-mass and high-mass inner planets, while Table \ref{table: occ rates r and m} gives the values. We find consistent occurrence rates between 1--10 and 10--20 M$_\oplus$ inner planets at the 0.2$\sigma$ level for metal-poor systems, 0.5$\sigma$ for metal-rich, and 0.8$\sigma$ for all metallicities. While these findings are tentative and certainly limited by small sample sizes, we intriguingly do not find increased significance of the difference between P(GG$|$1--10 M$_\oplus$) and P(GG$|$10--20 M$_\oplus$) when focused on just the metal-rich systems as compared to the total sample, in contrast to what we found when we looked at the radius valley divide. These findings mirror the tentative hints of enhanced inner-outer correlation in metal-rich systems with planets of high EMF and low bulk density with no obvious trend with the core mass.

\subsection{A cautionary tale}
\label{sec:caution}
These radius valley and mass occurrence rate comparisons have recently also been made in \citet{Chen2026}. In that paper, the authors similarly use our sample from \citet{BryanML2024}, taking the 109 systems that we labeled ``transit'' systems due to their discovery by the transit (rather than RV) method. Despite the same source sample and methodology adopted from \citet{BryanML2024}, \citet{Chen2026} present markedly different results from what we find in this paper. In sum, the authors find the occurrence of gas giants in metal-rich sub-Neptune systems to be significantly higher than that in metal-rich super-Earth systems. They also compare to the field star occurrence rate P(GG) and find consistent frequencies between P(GG) and P(GG$|$SE) but an enhancement for P(GG$|$SN). Finally, when \citet{Chen2026} take the subset of our 2024 sample that have small planets discovered using RVs, they find that there is a strong correlation between high-mass (10--20 M$_\oplus$) planets and outer gas giants in metal-rich systems in comparison to the field rate P(GG), and no correlation for the lower-mass (1--10 M$_\oplus$) sample.

While we identify three sources contributing to our different results, one has by far the largest impact: out-of-date system parameters whose effects on results are amplified due to small sample sizes. We briefly discuss this difference, as well as the other two contributing factors--- inaccurate completeness correction and different subsample definitions---in order of least to most impactful to the discrepancies with our results:

\begin{enumerate}
    \item Different subsample definitions: \citet{Chen2026} take the 109 ``transit'' system sample from \citet{BryanML2024} and divide it into three categories: systems with \textit{only} inner super-Earths (defined below the radius valley), systems with \textit{only} inner sub-Neptunes (above the radius valley), and systems with \textit{both} super-Earths and sub-Neptunes. Separate treatment of systems with both types of planets implicitly assumes a prior knowledge of the inner system completeness (e.g., that systems with only sub-Neptunes detected do not have unseen super-Earths). By comparison, our double-counting opts for a more conservative approach and applies subsample divisions that are agnostic to inner system completeness. 

    \item Inaccurate completeness corrections: As expressed in previous publications leveraging large sample sizes with heterogeneous datasets, it is important to correct sensitivity to distant gas giants on a system-by-system basis \citep{BryanML2019,BryanML2024,Bonomo2025,BryanML2025}, because individual system completeness can vary a lot given different number of datapoints, time baseline, precision, and observing cadence (see Section \ref{sec:injection recovery} for details). Especially with the small size of the subsamples, it is unlikely that these differences will effectively average out. For example, for our metal-rich sample of 56 systems divided across the radius valley, the super-Earth subset has $\sim$40$\%$ fewer ``missed'' planets than the sub-Neptune subset. It is therefore inaccurate to adopt an average completeness map for all systems as was done in \citet{Chen2026}.
    
    \item Out-of-date system parameters: We went through each metal-rich system with small planets (1--4 R$_\oplus$) in both the \citet{Chen2026} sample and the sample presented in this paper to identify discrepancies beyond subsample definitions. If we apply the same subsample definitions as those used by \citet{Chen2026} (SE, SN, mixed) to the sample presented in this paper, we get different subsample sizes and number of gas giants in each subsample. Mirroring Table 1 in \citet{Chen2026}, numbers we find versus theirs are:
    \begin{enumerate}
        \item Total SN systems: 26 vs. 23
        \item SN systems with GG: 9 vs. 9
        \item Total mixed systems: 16 vs. 18
        \item Mixed systems with GG: 4 vs. 3
        \item Total SE systems: 14 vs. 15
        \item SE systems with GG: 4 vs. 2
    \end{enumerate}
    These differences come from a handful of out-of-date system parameters used in \citet{Chen2026}. Three systems with no gas giants have small planets with radius measurements updated in the last $\sim$decade that move them across the radius valley, reclassifying these systems as follows: Mixed no GG $\rightarrow$ SN no GG, Mixed no GG $\rightarrow$ SN no GG, and SE no GG $\rightarrow$ Mixed no GG. In addition, there are two systems with outer gas giants that \citet{Chen2026} classified as having no outer giants, and one system with an outer GG that is missing from their sample. The two misclassified systems go from SE no GG $\rightarrow$ SE with GG, and Mixed no GG $\rightarrow$ Mixed GG. The missing system is 55 Cnc, which we include because 55 Cnc e is a small planet with a radius measurement, which was our criteria for inclusion.
\end{enumerate}

To illustrate the impact of just the out-of-date system parameters on the resulting occurrence rates, we compare raw occurrence rates (i.e. no completeness corrections) using the \citet{Chen2026} subsample definitions before and after system parameters have been correctly updated. We find:

\begin{enumerate}
    \item P(GG$|$SN, [Fe/H]$>$0) = 39.1$\%$ $\rightarrow$ 34.6$\%$
    \item P(GG$|$Mixed, [Fe/H]$>$0) = 16.7$\%$ $\rightarrow$ 25.0$\%$
    \item P(GG$|$SE, [Fe/H]$>$0) = 13.3$\%$ $\rightarrow$ 28.6$\%$
\end{enumerate}

While the frequency of gas giants in SN systems remains close, P(GG$|$SE) doubles, and the mixed sample occurrence goes up significantly. Even though the differences in the absolute number of systems in each subsample appear small, given small sample sizes the relative impact of these differences is significant, particularly because the changes primarily impact super-Earth systems. This example highlights the importance of careful planet characterization, especially in the small sample size regime.
    
We do not go into the occurrence rates calculated for small planets with measured masses, but simply note that the above three points apply to those calculations as well.

\subsection{Zeroing in on planets within the radius valley}

Thus far we have treated the radius valley as a binary divide, a boundary through the small planet period/radius/stellar mass parameter space above which lies the sub-Neptunes and below which super-Earths are found. However, this radius valley has a width, and is notably not empty. Here we investigate the occurrence of gas giants in systems with small planets \textit{within} the radius valley.

We define the valley width using the left panel in Figure 2 of \citet{HoCSK2023}. This figure illustrates the best fit radius valley boundary (black solid line) and width (black dashed lines) given just a period dependence, i.e. no stellar mass dependence. We opt to use this relation here because 1) the stellar mass dependence for our FGK star sample is minor and does not change out results in section 6.1, and 2) this Figure 2 in \citet{HoCSK2023} allows us to more clearly estimate the width of the radius valley as described below. 

The solid black line in \citet{HoCSK2023} Figure 2 shows the best-fit radius valley through radius-period space, defined in their equation 4:

\begin{equation}
    \log_{10}\left(\frac{R_p}{R_\oplus}\right) = -0.11\log_{10}\left(\frac{P}{\rm days}\right) + 0.37.
\end{equation}

The two dashed black lines parallel to the best fit indicate the upper and lower boundaries of the valley. We note by eye that the lower boundary passes through the y-axis at $\sim$2.0 R$_{\oplus}$. Since the best fit passes through at 10$^{0.37}$ = 2.34 R$_\oplus$, we take the upper y-intercept as 2.68 R$_\oplus$. Our in-valley planets (`VAL') are defined as those that lie within these boundaries of the radius valley and we find a total of 22 systems with such planets. We then redefine our SE samples to be planets that fall below the {\it lower bound} of the radius valley and our SN samples to be planets that lie above the {\it upper bound} of the radius valley.

The middle column of Figure \ref{fig:rad vally and mass cuts} shows these new three-subsample occurrence rates for all, high, and low metallicities. Since there are 22 systems total in the `VAL' subsample, occurrence rate posteriors are predictably broad. We will simply note here that the relative small offsets between P(GG$|$SE) and P(GG$|$SN) as a function of metallicity remain consistent with that reported for binary division (left column of Figure \ref{fig:rad vally and mass cuts}), and the occurrence of gas giants in systems with in-valley planets appears consistent with the SE/SN frequencies. The consistency of the valley population with SE/SN outer gas giant frequencies may suggest that the small planets within the radius valley are a continuation of the super-Earth to sub-Neptune populations rather than a distinct group of planets that followed special formation or evolutionary paths.

\begin{figure}
    \centering
    \includegraphics[width=0.93\linewidth]{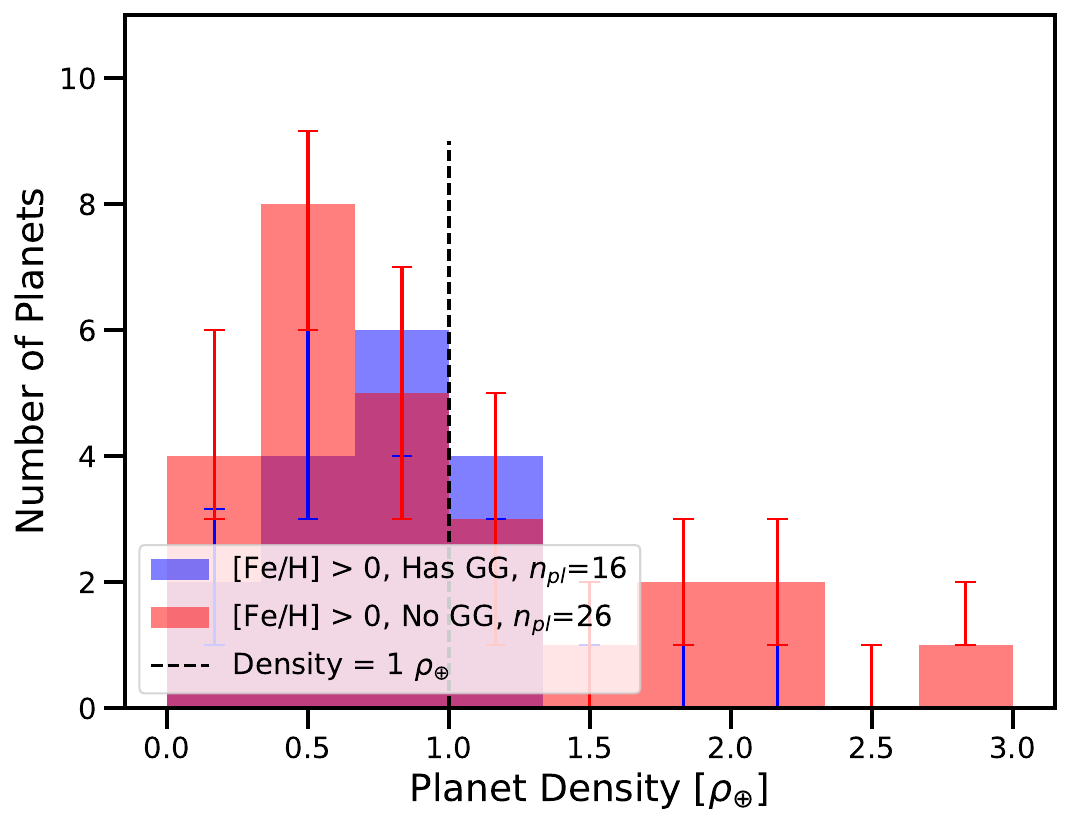}
    \includegraphics[width=0.93\linewidth]{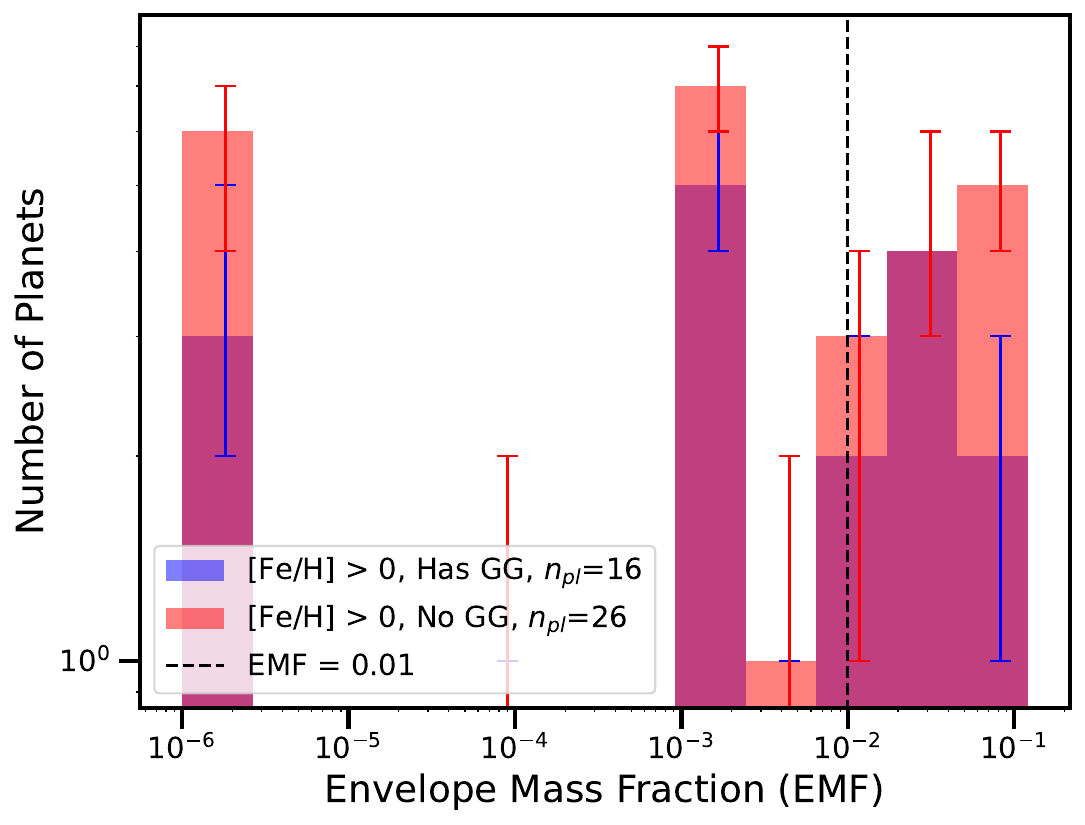}
    \includegraphics[width=0.93\linewidth]{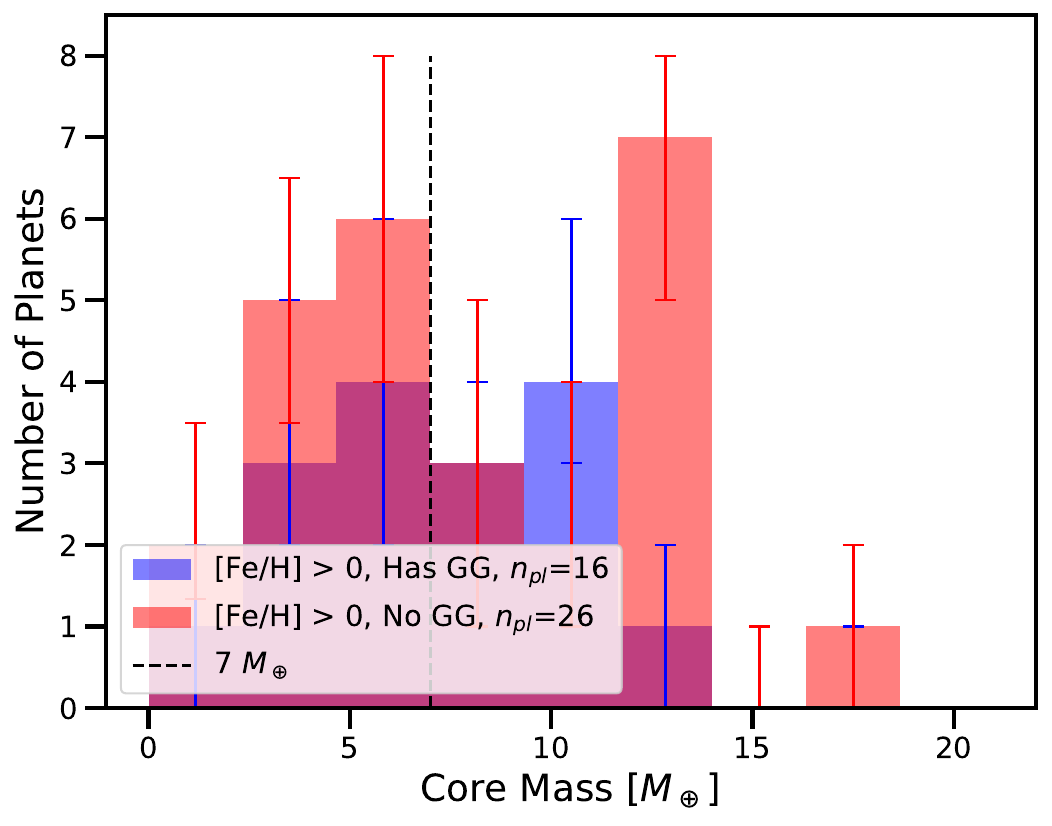}
    \caption{The same histograms as Figure~\ref{fig:pl hists} but now only including metal-rich systems. The gray vertical lines indicate the locations where we divided our sample to calculate conditional probabilities.
    }
    \label{fig:bias check}
\end{figure}

\section{Discussion}
\label{sec:discussion}
We consider the implications of these tentative findings in the broader context of planet formation. Previous work has demonstrated that a positive correlation between the inner small planets and outer gas giants is strongly mediated by host star metallicity \citep{ZhuW2018,BryanML2019,Zhu2024,BryanML2024,BryanML2025,Bonomo2025}. The frequency of gas giants in small planet systems is significantly higher than the frequency of gas giants around field stars when the host star is metal-rich ([Fe/H]$>$0.1), reflecting the critical role that metallicity, as a proxy for disk solid content, plays in shaping planetary architectures. Under the theory of pebble accretion, the efficiency at which the disk solids are converted to planetary objects is small $\lesssim$10\% as most solids drift past rather than being attracted to the planetary embryo \citep[e.g.,][]{ormel17,liu18,lin18}. It follows that when a disk has enough solids to form a gas giant on the outside, there will be enough material that drifts to the inner disk for small planets to also form \citep{chachan22,ChachanY2023}. Pebble accretion efficiency itself is independent of metallicity so the total amount of solid material delivered to the inner disk linearly increases with increasing initial disk metal content. 

When it comes to the question of whether these small planets grow into bona fide super-Earths (i.e. rocky cores with negligible atmospheres) or become sub-Neptunes with substantial gas envelope, the critical deciding factor is the mass of the rocky core, followed by the time at which the cores form before the disk gas dissipates away. The heavier the core is and the earlier it forms, the more gas it accretes \citep[e.g.,][]{Lee19}.\footnote{While the initial gas accretion physics expects a clear difference between the masses of super-Earths ($\lesssim$2$M_\oplus$) and sub-Neptunes ($\gtrsim$2$M_\oplus$) \citep{lee21,lee22}, more massive super-Earths can appear at short orbital periods ($\lesssim$30--50 days) following post-formation photoevaporative mass loss \citep[e.g.,][]{OwenJE2013}, which is observed to be the case.} Both conditions are expected to be met when there are more solids in the disk, which leads not only to more massive initial planetary embryos but also their faster mass growth by collisional mergers, resulting in more sub-Neptunes in metal-rich systems than rocky super-Earths \citep{DawsonRI2015}.  

Our tentative finding that puffy planets in metal-rich systems may be more likely to host an outer gas giant than denser rocky planets is consistent with the above story of how metallicity broadly controls system properties. More metal-rich systems have more solid content delivered to the inner disk, cores form faster, and accrete more gas before the disk dissipates. 

A question that remains unanswered in this picture is whether puffy sub-Neptunes appear due to the presence of the gas giant or simply a result of the protoplanetary disk solid content. We attempt to distinguish between these options with Figure~\ref{fig:bias check}, which shows the distribution of planet densities, core masses, and envelope mass fractions for only the metal-rich systems with and without gas giants. Overall, the distributions between the gas giant and non-gas giant systems appear consistent across inner planet properties. With the current sample size of metal-rich systems hosting inner small planets that have mass and radius measurements, we cannot conclude whether the gas giants in these systems play an active role in shaping small planet bulk composition---they may just be along for the ride. 

A complicating factor in this interpretation is the evolution of small planets due to photoevaporation. For planets forming close to their host stars, even if they initially form with significant envelopes, these can be stripped over time due to stellar irradiation \citep[e.g.,][]{OwenJE2013}, potentially contributing to why the gas giant occurrence rate appears invariant with the core mass of the inner planets. Furthermore, we do not find small planet orbital periods to be biased closer to or farther away from the host star, given the presence of an outer gas giant (Figure~\ref{fig:rad mass}). In fact, currently observed small planet population is complete only out to $\sim$30--50 days and their mass measurements are sensitive only out to $\sim$10--20 days. At these short periods, photoevaporative mass loss is expected to be fully operational \citep{Owen12,Owen17}. To control against post-formation mass loss, we need mass measurements beyond $\sim$50 days which is not available at this time.

\section{Conclusion} 
\label{sec:conclusion}
In this work we leveraged a sample of 43 systems hosting inner small planets with both measured masses and radii to explore whether or not gas giants impact small planet bulk densities. After accounting for differences in individual system sensitivities to outer gas giants, we confirmed that our subsample of small planet systems yields a consistent P(GG$|$ISP) in comparison to the larger small planet samples where we do not require the inner planets to have both a measured mass and radius for sample inclusion.

We assessed if gas giants correlate with inner small planet composition by calculating the probability of outer gas giants in systems with low/high planet densities, EMFs, and core masses. For our entire sample, there was no significant difference in gas giant occurrence across any of these properties. Breaking our sample into metal-rich and metal-poor systems, while we similarly see no difference in metal poor systems, we find a more intriguing {\it tentative} result for metal-rich systems. While probability distributions for the metal-rich subsample are broader due to smaller sample sizes, P(GG$|$ISP) increases for inner small planets of low bulk densities (compared to high bulk densities) and high EMF (compared to low EMF) with a larger significance than assessed with the entire sample. No noticeable difference in P(GG$|$ISP) was observed for the estimated core masses of the small planets. We find consistent trends using larger samples by relaxing our selection criteria to systems containing small planets with measured radii \textit{or} mass, and comparing gas giant occurrence rates in systems above/below the radius valley, and in systems with low/high mass small planets. 

Our results tentatively suggest that gas giants may be preferentially found in systems with puffy sub-Neptunes rather than rocky super-Earths, despite the limitations with small sample sizes. The same limitation of the sample size however precludes us from determining whether the puffy sub-Neptunes were shaped by gas giants or simply a reflection of the initial solid content of the protoplanetary disk. Either way, our tentative result is in alignment with the following three prior results: 1) the inner-outer planet correlation is stronger in metal-rich systems; 2) gas giants more easily nucleate in metal-heavy disks; 3) with low efficiency of pebble accretion, most solid mass drift to the inner disk so we expect more rapid core formation and therefore the creation of puffy sub-Neptunes when there are more solids to begin with in the disk. At first glance, this third point appears inconsistent with our result that P(GG$|$ISP) seems invariant with the estimated core mass of the inner planets. However, we note that this discrepancy is likely due to contamination by post-formation photoevaporation that strips initially gas-enveloped sub-Neptunes into rocky super-Earth planets.

To pursue the small planet/gas giant connection in this bulk composition space, we need larger sample sizes. In particular, we need more small planets with measured masses and radii around metal-rich stars. Ground-based RV follow-up efforts will be critical in this endeavor, as they provide not only the masses but also the baseline sensitivity needed to statistically interpret the occurrence of outer gas giants. In addition, adding small planets at longer orbital periods will enable tests of formation and gas accretion theories, disentangling from impacts of photoevaporation. We look forward to upcoming space telescopes such as PLATO for finding such planets.

\begin{acknowledgments}
JYT gratefully acknowledges support from NSERC through the NSERC USRA award. Additionally, we would like to thank the people who have kept the Endor cluster alive and running throughout this project. EJL was supported by NSF Research Grant 2509275.
\end{acknowledgments}

\vspace{5mm}

\appendix
\renewcommand{\thefigure}{A\arabic{figure}}
\setcounter{figure}{0}

\begin{table*}[t] 
\centering
\caption{Stellar properties for sample host stars}
\label{tab:systems}
\begin{tabular*}{\textwidth}{@{\extracolsep{\fill}} l c c c c c c c c}
\hline
Host & $M_\star$ & [Fe/H] & $L_\star$ & $N_p$ & GG & $N_{\rm obs}$ & Baseline (yr) & Ref. \\
\hline
55 Cnc      & $0.97^{+0.05}_{-0.05}$ & $0.38^{+0.06}_{-0.06}$ & $-0.20^{+0.01}_{-0.01}$ & 5 & Yes & 1584 & 30.6 & 1 \\
GJ 9827     & $0.61^{+0.01}_{-0.01}$ & $-0.26^{+0.09}_{-0.09}$ & $-1.00^{+0.04}_{-0.04}$ & 3 & No  & 128  & 10.8 & 2 \\
HD 136352   & $0.87^{+0.04}_{-0.04}$ & $-0.24^{+0.05}_{-0.05}$ & $0.03^{+0.03}_{-0.03}$ & 3 & No  & 458  & 19.3 & 3 \\
HD 207897   & $0.80^{+0.04}_{-0.04}$ & $-0.04^{+0.04}_{-0.04}$ & $-0.42^{+0.04}_{-0.04}$ & 1 & No  & 122  & 17.7 & 4 \\
KOI-142     & $0.99^{+0.02}_{-0.02}$ & $0.27^{+0.06}_{-0.06}$ & $-0.17^{+0.01}_{-0.01}$ & 3 & Yes & 47   & 9.0  & 5 \\
KOI-94      & $1.28^{+0.05}_{-0.05}$ & $0.02^{+0.01}_{-0.01}$ & $0.48^{+0.1}_{-0.1}$     & 4 & No  & 70   & 7.3  & 5 \\
Kepler-10   & $0.91^{+0.02}_{-0.02}$ & $-0.15^{+0.04}_{-0.04}$ & $0.05^{+0.01}_{-0.01}$ & 3 & No  & 291  & 11.0 & 2 \\
Kepler-100  & $1.09^{+0.03}_{-0.03}$ & $0.07^{+0.08}_{-0.05}$ & $0.41^{+0.02}_{-0.02}$ & 4 & No  & 112  & 13.4 & 5 \\
Kepler-102  & $0.80^{+0.02}_{-0.02}$ & $0.11^{+0.04}_{-0.04}$ & $-0.58^{+0.02}_{-0.01}$ & 5 & No  & 147  & 10.3 & 2 \\
Kepler-103  & $1.21^{+0.02}_{-0.03}$ & $0.16^{+0.04}_{-0.04}$ & $0.42^{+0.02}_{-0.02}$ & 2 & No  & 60   & 4.4  & 2 \\
Kepler-104  & $0.82^{+0.03}_{-0.03}$ & $-0.38^{+0.10}_{-0.10}$ & $0.09^{+0.02}_{-0.02}$ & 3 & No  & 44   & 11.3 & 5 \\
Kepler-106  & $0.96^{+0.03}_{-0.03}$ & $-0.12^{+0.11}_{-0.11}$ & $0.08^{+0.03}_{-0.03}$ & 4 & No  & 48   & 9.8  & 5 \\
Kepler-109  & $1.09^{+0.09}_{-0.08}$ & $-0.02^{+0.07}_{-0.07}$ & $0.35^{+0.03}_{-0.03}$ & 2 & No  & 66   & 10.0 & 2 \\
Kepler-11   & $0.99^{+0.03}_{-0.03}$ & $0.07^{+0.10}_{-0.10}$ & $0.03^{+0.01}_{-0.01}$ & 6 & No  & 31   & 11.6 & 5 \\
Kepler-113  & $0.79^{+0.02}_{-0.02}$ & $0.05^{+0.07}_{-0.07}$ & $-0.58^{+0.02}_{-0.02}$ & 2 & No  & 42   & 12.0 & 5 \\
Kepler-126  & $1.11^{+0.03}_{-0.03}$ & $-0.13^{+0.10}_{-0.10}$ & $0.45^{+0.02}_{-0.02}$ & 3 & No  & 35   & 5.3  & 5 \\
Kepler-129  & $1.24^{+0.04}_{-0.04}$ & $0.29^{+0.10}_{-0.10}$ & $0.46^{+0.02}_{-0.01}$ & 3 & Yes & 35   & 7.8  & 5 \\
Kepler-131  & $1.08^{+0.02}_{-0.02}$ & $0.12^{+0.07}_{-0.07}$ & $0.02^{+0.02}_{-0.01}$ & 2 & No  & 46   & 6.9  & 5 \\
Kepler-139  & $1.08^{+0.03}_{-0.03}$ & $0.27^{+0.10}_{-0.10}$ & $0.02^{+0.02}_{-0.02}$ & 5 & Yes & 38   & 11.8 & 5 \\
Kepler-1710 & $0.93^{+0.02}_{-0.02}$ & $-0.07^{+0.10}_{-0.10}$ & $-0.14^{+0.02}_{-0.02}$ & 1 & No  & 21   & 12.2 & 5 \\
Kepler-18   & $0.98^{+0.02}_{-0.02}$ & $0.20^{+0.10}_{-0.10}$ & $-0.03^{+0.04}_{-0.04}$ & 3 & No  & 25   & 8.8  & 5 \\
Kepler-19   & $0.90^{+0.02}_{-0.02}$ & $-0.13^{+0.06}_{-0.06}$ & $-0.16^{+0.01}_{-0.01}$ & 3 & No  & 73   & 9.3  & 5 \\
Kepler-20   & $0.93^{+0.05}_{-0.05}$ & $0.07^{+0.08}_{-0.08}$ & $-0.07^{+0.05}_{-0.05}$ & 6 & No  & 161  & 10.1 & 2 \\
Kepler-25   & $1.15^{+0.03}_{-0.03}$ & $-0.04^{+0.10}_{-0.10}$ & $0.47^{+0.02}_{-0.05}$ & 3 & No  & 99   & 11.3 & 5 \\
Kepler-36   & $1.03^{+0.04}_{-0.04}$ & $-0.18^{+0.04}_{-0.04}$ & $0.51^{+0.01}_{-0.02}$ & 2 & No  & 25   & 9.2  & 5 \\
Kepler-406  & $1.06^{+0.03}_{-0.03}$ & $0.18^{+0.07}_{-0.07}$ & $0.05^{+0.02}_{-0.02}$ & 2 & No  & 58   & 11.9 & 5 \\
Kepler-407  & $1.10^{+0.03}_{-0.03}$ & $0.33^{+0.07}_{-0.07}$ & $-0.04^{+0.01}_{-0.01}$ & 2 & Yes & 98   & 11.0 & 5 \\
Kepler-454  & $1.03^{+0.04}_{-0.03}$ & $0.32^{+0.08}_{-0.08}$ & $0.05^{+0.01}_{-0.02}$ & 3 & Yes & 147  & 12.1 & 2 \\
Kepler-48   & $0.92^{+0.02}_{-0.02}$ & $0.17^{+0.07}_{-0.07}$ & $-0.29^{+0.02}_{-0.04}$ & 5 & Yes & 59   & 12.8 & 5 \\
Kepler-50   & $1.15^{+0.04}_{-0.04}$ & $-0.04^{+0.10}_{-0.10}$ & $0.54^{+0.02}_{-0.02}$ & 2 & No  & 39   & 9.8  & 5 \\
Kepler-507  & $1.14^{+0.03}_{-0.03}$ & $0.16^{+0.10}_{-0.10}$ & $0.33^{+0.02}_{-0.02}$ & 1 & No  & 49   & 10.8 & 5 \\
Kepler-538  & $0.89^{+0.05}_{-0.04}$ & $-0.09^{+0.07}_{-0.07}$ & $-0.18^{+0.02}_{-0.01}$ & 1 & No  & 111  & 9.1  & 2 \\
Kepler-65   & $1.25^{+0.03}_{-0.03}$ & $0.17^{+0.06}_{-0.06}$ & $0.47^{+0.02}_{-0.02}$ & 4 & Yes & 79   & 11.1 & 5 \\
Kepler-68   & $1.06^{+0.02}_{-0.02}$ & $0.11^{+0.06}_{-0.06}$ & $0.23^{+0.02}_{-0.01}$ & 4 & Yes & 225  & 12.4 & 2 \\
Kepler-92   & $1.29^{+0.04}_{-0.04}$ & $0.14^{+0.01}_{-0.01}$ & $0.56^{+0.02}_{-0.01}$ & 3 & No  & 23   & 9.9  & 5 \\
Kepler-93   & $0.91^{+0.03}_{-0.03}$ & $-0.18^{+0.10}_{-0.10}$ & $-0.11^{+0.01}_{-0.01}$ & 2 & No  & 153  & 12.2 & 2 \\
Kepler-94   & $0.82^{+0.02}_{-0.02}$ & $0.34^{+0.07}_{-0.07}$ & $-0.61^{+0.02}_{-0.01}$ & 2 & Yes & 39   & 12.0 & 5 \\
Kepler-95   & $1.08^{+0.04}_{-0.04}$ & $0.30^{+0.10}_{-0.10}$ & $0.28^{+0.01}_{-0.02}$ & 1 & No  & 36   & 7.9  & 5 \\
Kepler-96   & $1.01^{+0.02}_{-0.02}$ & $0.04^{+0.07}_{-0.07}$ & $-0.05^{+0.02}_{-0.01}$ & 1 & No  & 55   & 10.9 & 5 \\
Kepler-97   & $0.90^{+0.03}_{-0.03}$ & $-0.20^{+0.07}_{-0.07}$ & $-0.03^{+0.02}_{-0.01}$ & 2 & No  & 31   & 8.0  & 5 \\
Kepler-98   & $1.00^{+0.02}_{-0.02}$ & $0.18^{+0.07}_{-0.07}$ & $-0.13^{+0.01}_{-0.02}$ & 1 & No  & 42   & 7.9  & 5 \\
Kepler-99   & $0.82^{+0.02}_{-0.02}$ & $0.18^{+0.07}_{-0.07}$ & $-0.59^{+0.03}_{-0.02}$ & 1 & No  & 45   & 7.0  & 5 \\
TOI-1736    & $1.08^{+0.04}_{-0.04}$ & $0.14^{+0.01}_{-0.01}$ & $0.27^{+0.03}_{-0.03}$ & 2 & Yes & 152  & 2.6  & 6 \\
\hline
\end{tabular*}%
\\[0.5ex] 
\parbox{\textwidth}{\centering \textbf{Note:} Gas Giants are defined as having masses 0.5--20 M$_{\rm Jup}$. \\ \textbf{References:} (1) \cite{RosenthalLJ2021}, (2) \cite{BonomoAS2023}, (3) \cite{KaneSR2020}, (4) \cite{HeidariN2022}, (5) \cite{WeissLM2024}, (6) \cite{MartioliE2023}}
\end{table*}

\begin{table*}[b]
\centering
\caption{Sample small planet properties}
\label{tab:SE}
\begin{tabular}{llccccccc}
\hline
Host & Planet & $R_p$ ($R_\oplus$) & $M_p\sin i$ ($M_\oplus$) & $a$ (AU) & Flux ($F_\oplus$) & $\rho$ (g cm$^{-3}$) & EMF$_{\rm st\,flux}$ & $M_{core}$ ($M_\oplus$) \\
\hline
55 Cnc & 55 Cnc e & $1.88^{+0.03}_{-0.03}$ & $7.99^{+0.33}_{-0.32}$ & $0.01544^{+0.00005}_{-0.00005}$ & $3504$ & $6.69^{+0.42}_{-0.41}$ & $1\times 10^{-6}$ & $7.99^{+0.33}_{-0.32}$ \\
GJ 9827 & GJ 9827 b & $1.44^{+0.07}_{-0.09}$ & $4.28^{+0.03}_{-0.04}$ & $0.0189^{+0.0004}_{-0.0004}$ & $1030$ & $7.91^{+1.15}_{-1.48}$ & $1\times 10^{-6}$ & $4.28^{+0.03}_{-0.04}$ \\
GJ 9827 & GJ 9827 c & $1.13^{+0.05}_{-0.07}$ & $1.86^{+0.39}_{-0.37}$ & $0.0395^{+0.0009}_{-0.0009}$ & $236$ & $7.11^{+1.76}_{-1.94}$ & $1\times 10^{-6}$ & $1.86^{+0.39}_{-0.37}$ \\
GJ 9827 & GJ 9827 d & $1.98^{+0.10}_{-0.11}$ & $3.02^{+0.57}_{-0.58}$ & $0.0530^{+0.0030}_{-0.0030}$ & $131$ & $2.15^{+0.52}_{-0.55}$ & $0.00616$ & $3.00^{+0.57}_{-0.58}$ \\
HD 136352 & HD 136352 b & $1.66^{+0.04}_{-0.04}$ & $4.72^{+0.42}_{-0.42}$ & $0.0964^{+0.0028}_{-0.0028}$ & $111$ & $5.65^{+0.67}_{-0.67}$ & $0.000100$ & $4.72^{+0.42}_{-0.42}$ \\
HD 136352 & HD 136352 c & $2.92^{+0.07}_{-0.07}$ & $11.2^{+0.6}_{-0.7}$ & $0.172^{+0.005}_{-0.005}$ & $34.9$ & $2.50^{+0.23}_{-0.24}$ & $0.0385$ & $10.8^{+0.6}_{-0.6}$ \\
HD 207897 & HD 207897 b & $2.50^{+0.08}_{-0.08}$ & $14.4^{+1.6}_{-1.6}$ & $0.116^{+0.002}_{-0.002}$ & $47.4$ & $5.08^{+0.74}_{-0.75}$ & $0.0122$ & $14.2^{+1.6}_{-1.6}$ \\
KOI-142 & KOI-142 b & $3.80^{+0.20}_{-0.20}$ & $11.4^{+5.1}_{-5.1}$ & $0.09582^{+0.00001}_{-0.00001}$ & $87.1$ & $1.15^{+0.54}_{-0.54}$ & $0.0970$ & $10.3^{+4.6}_{-4.6}$ \\
KOI-94 & KOI-94 b & $1.64^{+0.12}_{-0.12}$ & $5.75^{+3.80}_{-3.80}$ & $0.0498^{+0.0007}_{-0.0007}$ & $651$ & $7.22^{+5.04}_{-5.04}$ & $1\times 10^{-6}$ & $5.75^{+3.80}_{-3.80}$ \\
KOI-94 & KOI-94 c & $3.86^{+0.10}_{-0.10}$ & $6.20^{+5.59}_{-5.59}$ & $0.0986^{+0.0013}_{-0.0013}$ & $166$ & $0.593^{+0.536}_{-0.536}$ & $0.101$ & $5.58^{+5.02}_{-5.02}$ \\
Kepler-10 & Kepler-10 b & $1.49^{+0.04}_{-0.04}$ & $3.66^{+0.43}_{-0.43}$ & $0.0169^{+0.0001}_{-0.0001}$ & $3669$ & $6.11^{+0.85}_{-0.85}$ & $1\times 10^{-6}$ & $3.66^{+0.43}_{-0.43}$ \\
Kepler-10 & Kepler-10 c & $2.34^{+0.06}_{-0.06}$ & $12.1^{+2.6}_{-2.6}$ & $0.237^{+0.002}_{-0.002}$ & $18.5$ & $5.20^{+1.18}_{-1.18}$ & $0.00818$ & $12.0^{+2.6}_{-2.6}$ \\
Kepler-100 & Kepler-100 b & $1.34^{+0.13}_{-0.13}$ & $5.44^{+1.30}_{-1.30}$ & $0.0731^{+0.0001}_{-0.0001}$ & $278$ & $12.4^{+4.7}_{-4.7}$ & $1\times 10^{-6}$ & $5.44^{+1.30}_{-1.30}$ \\
Kepler-100 & Kepler-100 c & $2.34^{+0.08}_{-0.08}$ & $4.00^{+1.71}_{-1.71}$ & $0.1105^{+0.0001}_{-0.0001}$ & $121$ & $1.72^{+0.76}_{-0.76}$ & $0.0183$ & $3.93^{+1.68}_{-1.68}$ \\
Kepler-100 & Kepler-100 d & $1.54^{+0.23}_{-0.23}$ & $1.77^{+1.89}_{-1.89}$ & $0.2173^{+0.0001}_{-0.0001}$ & $31.4$ & $2.70^{+3.13}_{-3.13}$ & $0.00212$ & $1.77^{+1.89}_{-1.89}$ \\
Kepler-102 & Kepler-102 d & $1.34^{+0.09}_{-0.09}$ & $4.54^{+1.93}_{-1.93}$ & $0.0862^{+0.0008}_{-0.0008}$ & $77.5$ & $10.4^{+4.9}_{-4.9}$ & $1\times 10^{-6}$ & $4.54^{+1.93}_{-1.93}$ \\
Kepler-102 & Kepler-102 e & $2.41^{+0.14}_{-0.14}$ & $4.74^{+2.06}_{-2.06}$ & $0.116^{+0.001}_{-0.001}$ & $42.6$ & $1.86^{+0.87}_{-0.87}$ & $0.0203$ & $4.65^{+2.02}_{-2.02}$ \\
Kepler-103 & Kepler-103 b & $3.26^{+0.08}_{-0.08}$ & $13.9^{+6.9}_{-6.9}$ & $0.132^{+0.001}_{-0.001}$ & $87.4$ & $2.22^{+1.11}_{-1.11}$ & $0.0546$ & $13.1^{+6.5}_{-6.5}$ \\
Kepler-104 & Kepler-104 b & $2.38^{+0.06}_{-0.06}$ & $9.95^{+2.82}_{-2.82}$ & $0.0928^{+0.0001}_{-0.0001}$ & $127$ & $4.07^{+1.19}_{-1.19}$ & $0.0122$ & $9.82^{+2.78}_{-2.78}$ \\
Kepler-104 & Kepler-104 c & $2.36^{+0.08}_{-0.08}$ & $7.14^{+3.89}_{-3.89}$ & $0.1509^{+0.0001}_{-0.0001}$ & $48.0$ & $3.00^{+1.66}_{-1.66}$ & $0.0142$ & $7.04^{+3.84}_{-3.84}$ \\
Kepler-104 & Kepler-104 d & $2.63^{+0.14}_{-0.14}$ & $5.54^{+4.62}_{-4.62}$ & $0.2542^{+0.0001}_{-0.0001}$ & $16.9$ & $1.67^{+1.42}_{-1.42}$ & $0.0284$ & $5.38^{+4.49}_{-4.49}$ \\
Kepler-106 & Kepler-106 c & $2.39^{+0.07}_{-0.07}$ & $13.3^{+2.4}_{-2.4}$ & $0.1098^{+0.0001}_{-0.0001}$ & $90.1$ & $5.40^{+1.09}_{-1.09}$ & $0.00818$ & $13.2^{+2.4}_{-2.4}$ \\
Kepler-106 & Kepler-106 e & $2.62^{+0.17}_{-0.17}$ & $6.59^{+2.43}_{-2.43}$ & $0.2400^{+0.0001}_{-0.0001}$ & $18.9$ & $2.02^{+0.84}_{-0.84}$ & $0.0263$ & $6.42^{+2.37}_{-2.37}$ \\
Kepler-109 & Kepler-109 b & $2.33^{+0.07}_{-0.07}$ & $4.55^{+3.93}_{-3.93}$ & $0.0701^{+0.0019}_{-0.0019}$ & $288$ & $1.99^{+1.72}_{-1.72}$ & $0.0163$ & $4.48^{+3.87}_{-3.87}$ \\
Kepler-11 & Kepler-11 b & $1.92^{+0.06}_{-0.06}$ & $7.45^{+7.48}_{-7.48}$ & $0.0922^{+0.0010}_{-0.0010}$ & $121$ & $5.81^{+5.85}_{-5.85}$ & $0.00212$ & $7.43^{+7.46}_{-7.46}$ \\
Kepler-11 & Kepler-11 d & $3.38^{+0.10}_{-0.10}$ & $8.35^{+5.96}_{-5.96}$ & $0.156^{+0.001}_{-0.001}$ & $42.3$ & $1.19^{+0.86}_{-0.86}$ & $0.0708$ & $7.76^{+5.54}_{-5.54}$ \\
Kepler-113 & Kepler-113 b & $2.00^{+0.07}_{-0.07}$ & $7.53^{+3.37}_{-3.37}$ & $0.0512^{+0.0001}_{-0.0001}$ & $214$ & $5.18^{+2.38}_{-2.38}$ & $0.00212$ & $7.51^{+3.37}_{-3.37}$ \\
Kepler-113 & Kepler-113 c & $2.66^{+0.17}_{-0.17}$ & $1.05^{+3.28}_{-3.28}$ & $0.0779^{+0.0010}_{-0.0010}$ & $92.6$ & $0.307^{+0.961}_{-0.961}$ & $0.0223$ & $1.03^{+3.21}_{-3.21}$ \\
Kepler-126 & Kepler-126 b & $1.58^{+0.06}_{-0.06}$ & $4.12^{+7.12}_{-7.12}$ & $0.0971^{+0.0010}_{-0.0010}$ & $167$ & $5.77^{+10.00}_{-10.00}$ & $0.000100$ & $4.12^{+7.12}_{-7.12}$ \\
Kepler-126 & Kepler-126 c & $1.61^{+0.11}_{-0.11}$ & $1.28^{+7.03}_{-7.03}$ & $0.1584^{+0.0001}_{-0.0001}$ & $62.7$ & $1.70^{+9.34}_{-9.34}$ & $0.00212$ & $1.28^{+7.02}_{-7.02}$ \\
Kepler-129 & Kepler-129 b & $2.29^{+0.07}_{-0.07}$ & $10.2^{+6.2}_{-6.2}$ & $0.132^{+0.001}_{-0.001}$ & $91.0$ & $4.69^{+2.89}_{-2.89}$ & $0.00818$ & $10.1^{+6.2}_{-6.2}$ \\
Kepler-131 & Kepler-131 b & $2.06^{+0.05}_{-0.05}$ & $6.85^{+3.67}_{-3.67}$ & $0.1278^{+0.0001}_{-0.0001}$ & $62.6$ & $4.30^{+2.33}_{-2.33}$ & $0.00414$ & $6.82^{+3.65}_{-3.65}$ \\
Kepler-139 & Kepler-139 b & $2.38^{+0.07}_{-0.07}$ & $5.30^{+2.10}_{-2.10}$ & $0.126^{+0.001}_{-0.001}$ & $63.8$ & $2.18^{+0.88}_{-0.88}$ & $0.0183$ & $5.20^{+2.06}_{-2.06}$ \\
Kepler-139 & Kepler-139 d & $1.70^{+0.06}_{-0.06}$ & $4.65^{+2.00}_{-2.00}$ & $0.0756^{+0.0100}_{-0.0100}$ & $178$ & $5.27^{+2.34}_{-2.34}$ & $0.00212$ & $4.64^{+2.00}_{-2.00}$ \\
Kepler-1710 & Kepler-1710 b & $2.59^{+0.07}_{-0.07}$ & $3.49^{+5.77}_{-5.77}$ & $0.1158^{+0.0001}_{-0.0001}$ & $64.7$ & $1.11^{+1.84}_{-1.84}$ & $0.0284$ & $3.39^{+5.61}_{-5.61}$ \\
Kepler-18 & Kepler-18 b & $1.81^{+0.21}_{-0.21}$ & $12.8^{+3.9}_{-3.9}$ & $0.0449^{+0.0001}_{-0.0001}$ & $481$ & $11.9^{+5.5}_{-5.5}$ & $1\times 10^{-6}$ & $12.7^{+3.9}_{-3.9}$ \\
Kepler-19 & Kepler-19 b & $2.30^{+0.06}_{-0.06}$ & $7.40^{+2.11}_{-2.11}$ & $0.0836^{+0.0010}_{-0.0010}$ & $122$ & $3.35^{+0.99}_{-0.99}$ & $0.0122$ & $7.31^{+2.08}_{-2.08}$ \\
Kepler-20 & Kepler-20 b & $2.01^{+0.18}_{-0.18}$ & $9.51^{+1.45}_{-1.45}$ & $0.0459^{+0.0010}_{-0.0010}$ & $409$ & $6.49^{+2.03}_{-2.03}$ & $0.00212$ & $9.49^{+1.45}_{-1.45}$ \\
Kepler-20 & Kepler-20 c & $2.88^{+0.13}_{-0.13}$ & $13.6^{+2.2}_{-2.2}$ & $0.0940^{+0.0020}_{-0.0020}$ & $97.3$ & $3.13^{+0.68}_{-0.68}$ & $0.0324$ & $13.1^{+2.2}_{-2.2}$ \\
Kepler-20 & Kepler-20 d & $2.49^{+0.07}_{-0.07}$ & $12.9^{+4.2}_{-4.2}$ & $0.349^{+0.007}_{-0.007}$ & $7.06$ & $4.62^{+1.55}_{-1.55}$ & $0.0122$ & $12.8^{+4.2}_{-4.2}$ \\
Kepler-25 & Kepler-25 b & $2.70^{+0.06}_{-0.06}$ & $10.8^{+3.1}_{-3.1}$ & $0.0694^{+0.0010}_{-0.0010}$ & $316$ & $3.01^{+0.88}_{-0.88}$ & $0.0263$ & $10.5^{+3.0}_{-3.0}$ \\
Kepler-36 & Kepler-36 b & $1.49^{+0.06}_{-0.06}$ & $1.08^{+3.77}_{-3.77}$ & $0.114^{+0.001}_{-0.001}$ & $128$ & $1.82^{+6.34}_{-6.34}$ & $0.00212$ & $1.08^{+3.76}_{-3.76}$ \\
Kepler-36 & Kepler-36 c & $3.96^{+0.14}_{-0.14}$ & $16.3^{+8.7}_{-8.7}$ & $0.127^{+0.001}_{-0.001}$ & $103$ & $1.44^{+0.79}_{-0.79}$ & $0.103$ & $14.6^{+7.8}_{-7.8}$ \\
Kepler-406 & Kepler-406 b & $1.45^{+0.04}_{-0.04}$ & $5.33^{+1.76}_{-1.76}$ & $0.0360^{+0.0001}_{-0.0001}$ & $811$ & $9.63^{+3.29}_{-3.29}$ & $1\times 10^{-6}$ & $5.33^{+1.76}_{-1.76}$ \\
Kepler-407 & Kepler-407 b & $1.16^{+0.04}_{-0.04}$ & $1.67^{+0.82}_{-0.82}$ & $0.0153^{+0.0001}_{-0.0001}$ & $3826$ & $5.88^{+2.96}_{-2.96}$ & $1\times 10^{-6}$ & $1.67^{+0.82}_{-0.82}$ \\
Kepler-454 & Kepler-454 b & $1.84^{+0.06}_{-0.06}$ & $5.29^{+2.22}_{-2.22}$ & $0.0953^{+0.0010}_{-0.0010}$ & $116$ & $4.67^{+2.01}_{-2.01}$ & $0.00212$ & $5.27^{+2.21}_{-2.21}$ \\
\hline
\end{tabular}
\end{table*}

\begin{table*}[b] 
\centering
\parbox{\textwidth}{\centering \textbf{Table 3.} \textit{Continued.}}
\\[0.5ex]
\begin{tabular}{llccccccc}
\hline
Host & Planet & $R_p$ ($R_\oplus$) & $M_p\sin i$ ($M_\oplus$) & $a$ (AU) & Flux ($F_\oplus$) & $\rho$ (g cm$^{-3}$) & EMF$_{\rm st\,flux}$ & $M_{core}$ ($M_\oplus$) \\
\hline
Kepler-48 & Kepler-48 b & $1.85^{+0.09}_{-0.09}$ & $6.61^{+2.83}_{-2.83}$ & $0.0539^{+0.0010}_{-0.0010}$ & $258$ & $5.71^{+2.60}_{-2.60}$ & $0.00212$ & $6.59^{+2.83}_{-2.83}$ \\
Kepler-48 & Kepler-48 c & $2.56^{+0.07}_{-0.07}$ & $11.1^{+3.3}_{-3.3}$ & $0.0863^{+0.0010}_{-0.0010}$ & $101$ & $3.65^{+1.12}_{-1.12}$ & $0.0183$ & $10.9^{+3.2}_{-3.2}$ \\
Kepler-48 & Kepler-48 d & $1.98^{+0.07}_{-0.07}$ & $8.89^{+5.65}_{-5.65}$ & $0.233^{+0.001}_{-0.001}$ & $13.8$ & $6.32^{+4.06}_{-4.06}$ & $0.00212$ & $8.87^{+5.64}_{-5.64}$ \\
Kepler-50 & Kepler-50 b & $1.54^{+0.05}_{-0.05}$ & $4.02^{+8.93}_{-8.93}$ & $0.0808^{+0.0010}_{-0.0010}$ & $262$ & $6.05^{+13.45}_{-13.45}$ & $1\times 10^{-6}$ & $4.02^{+8.93}_{-8.93}$ \\
Kepler-50 & Kepler-50 c & $1.82^{+0.18}_{-0.18}$ & $10.2^{+10.2}_{-10.2}$ & $0.0913^{+0.0100}_{-0.0100}$ & $205$ & $9.26^{+9.69}_{-9.69}$ & $1\times 10^{-6}$ & $10.2^{+10.2}_{-10.2}$ \\
Kepler-507 & Kepler-507 b & $1.28^{+0.04}_{-0.04}$ & $5.69^{+1.47}_{-1.47}$ & $0.0478^{+0.0010}_{-0.0010}$ & $607$ & $14.8^{+4.1}_{-4.1}$ & $1\times 10^{-6}$ & $5.69^{+1.47}_{-1.47}$ \\
Kepler-538 & Kepler-538 b & $2.16^{+0.05}_{-0.05}$ & $10.5^{+4.9}_{-4.9}$ & $0.353^{+0.010}_{-0.010}$ & $6.70$ & $5.75^{+2.73}_{-2.73}$ & $0.00414$ & $10.4^{+4.9}_{-4.9}$ \\
Kepler-65 & Kepler-65 b & $1.52^{+0.09}_{-0.09}$ & $2.45^{+2.39}_{-2.39}$ & $0.0352^{+0.0010}_{-0.0010}$ & $1290$ & $3.85^{+3.81}_{-3.81}$ & $1\times 10^{-6}$ & $2.45^{+2.39}_{-2.39}$ \\
Kepler-65 & Kepler-65 c & $2.58^{+0.06}_{-0.06}$ & $4.98^{+4.01}_{-4.01}$ & $0.0685^{+0.0020}_{-0.0020}$ & $340$ & $1.59^{+1.29}_{-1.29}$ & $0.0284$ & $4.84^{+3.90}_{-3.90}$ \\
Kepler-65 & Kepler-65 d & $1.78^{+0.11}_{-0.11}$ & $4.24^{+3.93}_{-3.93}$ & $0.0852^{+0.0100}_{-0.0100}$ & $220$ & $4.15^{+3.92}_{-3.92}$ & $0.00212$ & $4.23^{+3.92}_{-3.92}$ \\
Kepler-68 & Kepler-68 b & $2.31^{+0.05}_{-0.05}$ & $8.44^{+1.50}_{-1.50}$ & $0.0611^{+0.0010}_{-0.0010}$ & $326$ & $3.78^{+0.72}_{-0.72}$ & $0.0102$ & $8.35^{+1.48}_{-1.48}$ \\
Kepler-92 & Kepler-92 b & $3.70^{+0.12}_{-0.12}$ & $19.4^{+8.4}_{-8.4}$ & $0.122^{+0.010}_{-0.010}$ & $117$ & $2.10^{+0.93}_{-0.93}$ & $0.0788$ & $17.8^{+7.7}_{-7.7}$ \\
Kepler-92 & Kepler-92 c & $2.45^{+0.07}_{-0.07}$ & $12.7^{+10.3}_{-10.3}$ & $0.190^{+0.001}_{-0.001}$ & $48.2$ & $4.75^{+3.86}_{-3.86}$ & $0.0102$ & $12.6^{+10.1}_{-10.1}$ \\
Kepler-92 & Kepler-92 d & $2.06^{+0.07}_{-0.07}$ & $11.9^{+11.3}_{-11.3}$ & $0.287^{+0.001}_{-0.001}$ & $21.3$ & $7.46^{+7.14}_{-7.14}$ & $0.00212$ & $11.8^{+11.3}_{-11.3}$ \\
Kepler-93 & Kepler-93 b & $1.63^{+0.06}_{-0.06}$ & $3.60^{+0.57}_{-0.57}$ & $0.0527^{+0.0006}_{-0.0006}$ & $324$ & $4.60^{+0.90}_{-0.90}$ & $0.00212$ & $3.59^{+0.57}_{-0.57}$ \\
Kepler-94 & Kepler-94 b & $3.04^{+0.12}_{-0.12}$ & $11.0^{+2.6}_{-2.6}$ & $0.0337^{+0.0010}_{-0.0010}$ & $476$ & $2.15^{+0.57}_{-0.57}$ & $0.0445$ & $10.5^{+2.5}_{-2.5}$ \\
Kepler-95 & Kepler-95 b & $3.12^{+0.09}_{-0.09}$ & $10.2^{+3.4}_{-3.4}$ & $0.102^{+0.001}_{-0.001}$ & $127$ & $1.87^{+0.64}_{-0.64}$ & $0.0506$ & $9.70^{+3.25}_{-3.25}$ \\
Kepler-96 & Kepler-96 b & $2.37^{+0.06}_{-0.06}$ & $13.0^{+4.7}_{-4.7}$ & $0.126^{+0.001}_{-0.001}$ & $59.8$ & $5.35^{+1.96}_{-1.96}$ & $0.00818$ & $12.9^{+4.6}_{-4.6}$ \\
Kepler-97 & Kepler-97 b & $1.62^{+0.09}_{-0.09}$ & $3.41^{+2.52}_{-2.52}$ & $0.0356^{+0.0010}_{-0.0010}$ & $769$ & $4.40^{+3.33}_{-3.33}$ & $1\times 10^{-6}$ & $3.41^{+2.52}_{-2.52}$ \\
Kepler-98 & Kepler-98 b & $1.87^{+0.13}_{-0.13}$ & $3.07^{+2.08}_{-2.08}$ & $0.0261^{+0.0010}_{-0.0010}$ & $1283$ & $2.59^{+1.83}_{-1.83}$ & $0.00212$ & $3.06^{+2.08}_{-2.08}$ \\
Kepler-99 & Kepler-99 b & $1.81^{+0.14}_{-0.14}$ & $3.81^{+2.06}_{-2.06}$ & $0.0507^{+0.0010}_{-0.0010}$ & $217$ & $3.53^{+2.07}_{-2.07}$ & $0.00212$ & $3.80^{+2.06}_{-2.06}$ \\
TOI-1736 & TOI-1736 b & $3.18^{+0.09}_{-0.12}$ & $12.3^{+1.4}_{-1.5}$ & $0.0730^{+0.0010}_{-0.0010}$ & $241$ & $2.11^{+0.30}_{-0.35}$ & $0.0506$ & $11.7^{+1.3}_{-1.4}$ \\
\hline
\end{tabular}
\\[0.5ex] 
\parbox{\textwidth}{\centering \textbf{Note:} Small planets are defined as having mass 1--20 M$_{\oplus}$ and radius 1--4 R$_{\oplus}$.}
\end{table*}

\begin{table*}[t]
\centering
\caption{Sample gas giant properties}
\label{tab:GG}
\begin{tabular*}{\textwidth}{@{\extracolsep{\fill}} lccccl}
\hline
Planet & $M_p \sin i$ ($M_J$) & $a$ (AU) & $M_\star$ ($M_\odot$) & [Fe/H] & Type \\
\hline
KOI-142 c   & $0.653^{+0.02}_{-0.02}$ & $0.1541^{+0.000001}_{-0.000001}$ & $0.99^{+0.02}_{-0.02}$ & $0.27^{+0.06}_{-0.06}$ & Cold \\
Kepler-88 d & $3.15^{+0.2}_{-0.2}$   & $2.47^{+0.01}_{-0.01}$           & $0.99^{+0.02}_{-0.02}$ & $0.27^{+0.06}_{-0.06}$ & Warm \\
Kepler-129 d& $6.34^{+0.1}_{-0.1}$   & $3.26^{+0.1}_{-0.1}$             & $1.24^{+0.04}_{-0.04}$ & $0.29^{+0.1}_{-0.1}$ & Warm  \\
Kepler-139 e& $1.34^{+0.1}_{-0.1}$   & $3.24^{+0.01}_{-0.01}$           & $1.08^{+0.03}_{-0.03}$ & $0.27^{+0.1}_{-0.1}$ & Warm  \\
Kepler-407 c& $11.1^{+0.05}_{-0.05}$ & $3.27^{+0.001}_{-0.001}$         & $1.10^{+0.03}_{-0.03}$ & $0.33^{+0.07}_{-0.07}$ & Warm  \\
Kepler-454 c& $4.58^{+0.05}_{-0.05}$ & $1.30^{+17}_{-17}$               & $1.03^{+0.04}_{-0.03}$ & $0.32^{+0.08}_{-0.08}$ & Warm  \\
Kepler-454 d& $2.31^{+0.4}_{-0.4}$   & $5.68^{+0.3}_{-0.3}$             & $1.03^{+0.04}_{-0.03}$ & $0.32^{+0.08}_{-0.08}$ & Warm  \\
Kepler-48 e & $2.16^{+0.06}_{-0.06}$ & $1.898^{+0.001}_{-0.001}$        & $0.92^{+0.02}_{-0.02}$ & $0.17^{+0.07}_{-0.07}$ & Warm  \\
Kepler-48 f & $0.94^{+0.3}_{-0.3}$   & $5.719^{+0.001}_{-0.001}$        & $0.92^{+0.02}_{-0.02}$ & $0.17^{+0.07}_{-0.07}$ & Warm  \\
Kepler-65 e & $0.68^{+0.07}_{-0.07}$ & $0.854^{+0.01}_{-0.01}$          & $1.25^{+0.03}_{-0.03}$ & $0.17^{+0.06}_{-0.06}$ & Cold  \\
Kepler-68 d & $0.742^{+0.03}_{-0.03}$& $1.46^{+0.01}_{-0.01}$           & $1.06^{+0.02}_{-0.02}$ & $0.11^{+0.06}_{-0.06}$ & Warm  \\
Kepler-68 e & $0.52^{+0.1}_{-0.1}$   & $5.38^{+0.1}_{-0.1}$             & $1.06^{+0.02}_{-0.02}$ & $0.11^{+0.06}_{-0.06}$ & Warm  \\
Kepler-94 c & $8.88^{+0.1}_{-0.1}$   & $1.597^{+0.0001}_{-0.0001}$      & $0.82^{+0.02}_{-0.02}$ & $0.34^{+0.07}_{-0.07}$ & Warm  \\
TOI-1736 c  & $8.81^{+2}_{-0.6}$     & $1.38^{+0.02}_{-0.02}$           & $1.08^{+0.04}_{-0.04}$ & $0.14^{+0.01}_{-0.01}$ & Warm  \\
55 Cnc b    & $0.840^{+0.03}_{-0.03}$& $0.116^{+0.002}_{-0.002}$        & $0.97^{+0.05}_{-0.05}$ & $0.38^{+0.06}_{-0.06}$ & Cold  \\
55 Cnc d    & $2.86^{+0.2}_{-0.2}$   & $5.54^{+0.1}_{-0.1}$             & $0.97^{+0.05}_{-0.05}$ & $0.38^{+0.06}_{-0.06}$ & Warm  \\
\hline
\end{tabular*}
\\[0.5ex] 
\parbox{\textwidth}{\centering \textbf{Note:} Gas giants have masses 0.5--20 M$_{\rm Jup}$. Cold Jupiters orbit 1--10AU, while warm Jupiters orbit 0.1--1AU.}
\end{table*}

\clearpage
\newpage

\bibliography{gg_se_corr}{}
\bibliographystyle{aasjournalv7}

\end{document}